\documentclass[twocolumn,prd,groupedaddress,nopacs,nofootinbib,floats]{revtex4}
\pdfoutput=1
\usepackage{graphicx}

\def\be{\begin{equation}}
\def\ee{\end{equation}}
\newcommand{\bea}{\begin{eqnarray}}
\newcommand{\eea}{\end{eqnarray}}
\newcommand{\hu}{\text{km\,s}^{-1}\text{Mpc}^{-1}}

\newcommand{\omin}{\Omega_m^{(\text{in})}}

\newcommand{\omout}{\Omega_m^{(\text{out})}}

\renewcommand{\in}{{(\text{in})}}
\newcommand{\out}{{(\text{out})}}

\newcommand{\dec}{{*}}
\newcommand{\eq}{{\text{eq}}}
\newcommand{\eff}{{\text{eff}}}

\begin{document}

\title{
The Cosmic Microwave Background in an Inhomogeneous Universe\\
\it --   why void models of dark energy are only weakly constrained by the CMB --
}

\author{Chris Clarkson$^1$ and Marco Regis$^{1,2}$}

\affiliation{$^1$Centre for Astrophysics, Cosmology \& Gravity, and, Department of Mathematics \& Applied Mathematics, University of Cape Town, Rondebosch 7701, Cape Town, South Africa\\ $^2$Centre for High Performance Computing, 15 Lower Hope St, Rosebank, Cape Town, South Africa}

\begin{abstract}

The dimming of Type Ia supernovae could be the result of Hubble-scale inhomogeneity in the matter and spatial curvature, rather than signaling the presence of a dark energy component. A key challenge for such models is to fit the detailed spectrum of the cosmic microwave background (CMB). We present a detailed discussion of the small-scale CMB in an inhomogeneous universe, focusing on spherically symmetric `void' models. We allow for the dynamical effects of radiation while analyzing the problem, in contrast to other work which inadvertently fine tunes its spatial profile. This is a surprisingly important effect and we reach substantially different conclusions. Models which are open at CMB distances fit the CMB power spectrum without fine tuning; these models also fit the supernovae and local Hubble rate data which favour a high expansion rate.  Asymptotically flat models may fit the CMB, but require some extra assumptions. We argue that a full treatment of the radiation in these models is necessary if we are to understand the correct constraints from the CMB, as well as other observations which rely on it, such as spectral distortions of the black body spectrum, the kinematic Sunyaev-Zeldovich effect or the Baryon Acoustic Oscillations.

\end{abstract}

\maketitle


\section{Introduction}

The problem of understanding the physical origin and value of the cosmological constant is leading us to reconsider some of the foundational aspects of cosmological model building more carefully~\cite{CM}. In particular, it is an important fact that, at the moment, the spatial homogeneity of the universe on Gpc scales exists by assumption, and is not yet an observationally proven fact. Given this uncertainty, so-called void models can explain the observed distance modulus utilizing a spatially varying energy density, Hubble rate and curvature on Gpc scales, without any unusual physical fields at late times~\cite{MT2,MHE,zehavi,PascualSanchez:1999zr,celerier1,Tomita:2000jj,tomita2001,hellaby,IKN,Mof1,Mof2,AAG,VFW,Alnes:2006pf,CR,Celerier:2006gy,garfinkle,chung&romano,BMN,EM,alnes,Romano:2007zz,celerier3,conley,sarkar,mattsson,ABNV,bolejko,enqvist,Caldwell:2007yu,CBL,UCE,zibin,YKN,ishak,CFL,GarciaBellido:2008gd,huntsarkar,ZMS,gbh1,CFZ,BW,CCF,CBKH,Tomita:2009wz,tomita09,FLSC,mortsell,Mof3,Romano:2009mr,Garfinkle:2009uf,Romano:2009qx,Kolb:2009hn,Romano:2009ej,peter,ages,RC,Yoo:2010qy,BNV,MZS,Yoo:2010ad}\footnote{We have in mind the `very big void models' which vary gently over Gpc scales (e.g.,~\cite{MT2,AAG,FLSC}). The density profile reaches full width at half maximum around the Hubble scale. Other `void models'~-- or Hubble bubble models~-- for dark energy suggest that because very big voids of 100's Mpc across are observed to exist, we could live in the centre of one of these, giving the necessary jump in the distance modulus to fit the SNIa~-- see, e.g.,~\cite{zehavi,BMN,conley,huntsarkar,ABNV}. These can be tested with sufficiently many SNIa~\cite{CFL}, and are often discussed in a perturbed FLRW context. In essence, our analysis here applies to these models too, but we don't consider these specifically (though see Sec.~\ref{cmbcalc}). }. Although introduced before the SNIa data~\cite{MT2}, they are currently attracting attention as, in some respects, one of the most conservative explanations for the dark energy problem. These models require an as yet unknown mechanism for their formation, probably requiring something unusual at the start of inflation to create a model close to spherically symmetric on Hubble scales. They also are in (dire) need of an explanation for the anti-Copernican fine tuning that exists: we have to be within tens of Mpc of the centre of spherical symmetry of the background~\cite{Alnes:2006pf,alnes,mortsell}, which implies a coincidence of, roughly, (40 Mpc/15 Gpc)$^3\sim 10^{-8}$ (though see Sec.~\ref{sec:disc}). Nevertheless, given that even worse temporal fine-tuning exists in our present understanding of the value of the cosmological constant and the coincidence problem, these models should be taken seriously despite this sinister drawback.

Observationally, it is very hard to disentangle time evolution from radial variation for a central observer (which we consider here). To make matters worse, our current lack of a plausible formation mechanism means that observationally, at this stage, we should think in terms of directly reconstructing the void, rather than constraining a model parametrically. In this sense,
we have to consider the physical conditions for the centre of the void and asymptotically as essentially independent to be determined observationally (perhaps  assuming that all quantities have the same spatial profile). A common theme is to imagine a void in an Einstein-de Sitter model to make things simpler (e.g.,~\cite{ZMS}); but they are so big as to be impossible to create from scale-invariant Gaussian initial conditions. There seems no real reason to be restricted to asymptotically flat models if we know slow-roll inflation has to be reworked anyway. 

What observations can reconstruct the void? SNIa and other local observations give us a handle on the curvature locally which tells us the Hubble rate and total matter density locally, as well as their spatial profiles out to a redshift of $\mathcal{O}(1)$. CMB observations help constrain the void profile out to large distances, because the CMB gives us a very precise measurement of the area distance at $z\sim1100$. Moreover, as we discuss in this paper, the CMB tells us the baryon fraction and the baryon-to-photon ratio at comoving scales $\sim13\,$Gpc away from us. To fully constrain the possible degrees of freedom, we need other observations to tell us how the baryon fraction, the radiation energy density and the primordial power spectrum vary with radius.

The physics of decoupling and line-of-sight effects contribute differently to the CMB, and have different dependency on the cosmological model. In sophisticated inhomogeneous models both pre- and post-decoupling effects will play a role, but Hubble-scale void models allow an important simplification for calculating the moderate to high $\ell$ part of the CMB~\cite{ZMS,CFZ} (see also~\cite{Vonlanthen:2010cd}). 
The comoving scale of the voids which closely mimic the $\Lambda$CDM distance modulus are typically $\mathcal{O}(\mbox{Gpc})$. The physical size of the sound horizon, which sets the largest scale seen in the pre-decoupling part of the power spectrum, is around $150\,$Mpc in comoving units.
This implies that in any causally connected patch of the Universe prior to decoupling, the density gradient is very small. Furthermore, the comoving radius of decoupling is larger than $10\,$Gpc, on which scale the gradient of the void profile is small anyway (by assumption of course - it could be set to zero). For example, at decoupling the total fractional difference in energy density  between the centre of the void and the asymptotic region is around 10\%~\cite{RC}; hence, across a causal patch we expect a maximum 1\% change in the energy density in the radial direction, and much less at the radius of the CMB that we observe for a Gaussian profile. This suggests that before decoupling on small scales we can model the universe in disconnected FLRW shells at different radii, with the one of interest located at the distance where we see the CMB. This can be calculated using standard FLRW codes, but with the line-of-sight parts corrected for~\cite{ZMS,CFZ,Vonlanthen:2010cd}.

For line-of-sight effects, we need to use the full void model. These come in two forms. The simplest effect is via the background dynamics, which affects the area distance to the CMB, somewhat similar to a simple dark energy model. This is the important effect for the small-scale CMB. The more complicated effect is on the largest scales through the Integrated Sachs-Wolfe effect (see~\cite{Tomita:2009wz} for the general formulas in LTB), which we don't consider here, since it depends on the detailed evolution of perturbations during the curvature era which are not yet understood in the context of inhomogeneous models (although some progress has been made~\cite{zibin,CCF,peter}).

The CMB has been considered in void models before using these approximations. Several papers have computed how the position of the peaks constrain the model~\cite{AAG,Alnes:2006pf,BW,Yoo:2010qy}. In~\cite{CFZ}, the full spectrum on small scales (say, $\ell\gtrsim100$) was shown to fit essentially perfectly, following from earlier work by \cite{ZMS}. These papers found that, although the CMB can be accommodated, it required either a very low Hubble rate at the centre or very high curvature for the simplest asymptotically flat void models, so ruling them out. Models which were not asymptotically flat also had problems~\cite{CFZ}. However, in~\cite{CFZ}, it was shown that a varying bang time could be introduced to make the CMB fit for sensible central values of measured parameters. 

More recent work~\cite{BNV,MZS}, released simultaneously with the first version of this paper, used these results and combined them with other data sets to constrain the models~\cite{BNV} or even rule out void models altogether~\cite{MZS} (when the bang time is homogeneous). The main constraint comes from the need of a small value of $H_0$ for compatibility with the CMB. 
These analyses both used methods for fitting the CMB which, we shall argue, over-constrain the models by effectively fine-tuning the radiation profile unnecessarily. 

A common theme in  these papers was that the influence of radiation was largely neglected and treated as a test field in an LTB void model when performing the matching to an early-time FLRW solution. This seems reasonable because radiation only contributes about 10\% of the energy density at last scattering. Furthermore, the area distance-redshift relation only varies by at most a few percent in FLRW models at $z\sim1100$, depending on whether radiation is included or not. It seems sensible that any errors induced by this approximation would lead to percent-level errors in any subsequent parameter estimation. 

We shall argue here that this is not the case. By including the effects of radiation in the dynamics of the spacetime, we find that the CMB constraints are considerably less restrictive, even for a high value of the Hubble rate at the centre (i.e., $h\sim0.7$). We also find that asymptotically open models are preferred, not closed ones. The effects of radiation are important, as first suggested in~\cite{RC}, an analysis we expand upon in detail here. There are several (interconnected) ways to see this: one is simply that the extra degree of freedom from including inhomogeneous radiation absorbs a constraint which, we argue, is artificial in pure dust LTB (as we describe in more details at the end of Sec.~\ref{cmbth}); another is that the dynamics of radiation at early times imprints itself in the dynamics at late times (see Appendix~\ref{apprad}); a third is that CMB observations cannot determine, on their own, the interior of our past lightcone, which appears as a repercussion in other analyses. 

In previous analyses an LTB relation was used in going from scale factor to redshift, a function $a(z)$. In LTB, this differs from the FLRW relation $a(z)=1/(1+z)$ by tipically a few percent out to $z\sim1100$. A direct consequence of this relation is that the CMB temperature today, on a surface of constant time, varies by at most a few percent across the void for a monotonic profile. Consequently, the variation in the radiation density must be homogeneous at the sub-percent level. In effect, this is the condition which requires a very low Hubble rate at the centre. Does this make sense? Intuitively it does not because the difference in the Hubble rate inside and outside the void at late times varies typically by tens of percent, and the density by even more than this~-- factors of 5 or more are not unreasonable. So, inside the void, an observer is expanding away from the CMB significantly faster than an observer outside. Surely they would measure a significantly cooler CMB than an equivalent observer outside~-- by much more than the percent level?

More concretely, consider a void much larger than the Hubble scale today, so that we can safely use FLRW results inside the void, and asymptotically far from the void. If the model has homogeneous baryon fraction and baryon-to-photon ratio (say, as an example to set the radiation to matter ratio), then the CMB will be emitted at constant temperature, and constant Hubble rate. Today, the temperature in any region where we can neglect the effect of inhomogeneity must locally satisfy $T_0^3\propto\Omega_m h^2$; depending on the density of the void, this can vary by a large fraction from inside to outside. 

As a final way to see the freedom inherent in the radiation, consider that we must be able to specify a spherically symmetric model as a Cauchy problem in the following manner. We have the freedom to specify both the radiation density and matter density as independent arbitrary profiles on a hypersurface of constant time \emph{today}, and evolve them backwards to the last scattering surface. The Hubble rate today is also a free function, but this can be specified iteratively to make the bang time homogeneous if desired. This is a perfectly feasible way to make a model (in principle), and demonstrates that the temperature today is not forced to be near-homogeneous~-- it can be anything, and is independent of the matter profile. As far as we are concerned, if these profiles are chosen such that the temperature of the CMB at the centre is as measured (with all other central parameters chosen as desired), and parameters asymptotically have the correct baryon fraction and baryon-photon ratio (which fixes the CMB peaks), then we have a candidate model for the CMB; as we discuss, only the area distance provides an additional constraint (to place the peaks in the right place). Once these are fixed the model can be evolved backwards to find out the interior of past lightcone of the central observer~-- it is not fixed a priori.  This argument is really just stating the obvious: that observations of the CMB \emph{cannot}, on their own, tell us about conditions in the interior of our past lightcone.  \footnote{There is an important exception to this statement given by perturbations in the observed CMB spectrum induced by microwave photons rescattered in clusters or inter-cluster medium which can in principle (together with baryon acoustic oscillation measurements) probe the inhomogeneity profile in the interior of our past lightcone, as we will discuss in Sec.~IV.}

To use the small-scale CMB to constrain void models in full generality actually requires a two-fluid radiation-plus-matter solution of the field equations in which the two fluid are non-comoving in the radial direction (in the background). (Strictly speaking, we need a three-fluid description because the dark matter and baryons decouple well before last scattering.) As we shall discuss, this is quite a challenging problem in general. 
However, the physics of decoupling (which we assume can be described as in FLRW) unambiguously fixes the baryon fraction and baryon-photon ratio at last scattering.
Then combining these constraints with bounds on the area distance (which encodes line-of-sight effects), the late time matter void profile and the radiation profile are constrained. 
For the latter, we shall argue, by analyzing the field equations using the approximation that the radial velocity between the matter and the radiation is small, that models can be constructed which can fit the CMB, alongside other observations (SNIa and $H_0$). However, given the uncertainty in the theoretical description, a full likelihood analysis is premature, and not considered here.

This paper is organized as follows. First we review the CMB in FLRW, extracting the important parts for our analysis. This will show how to disentangle effects associated to physics of decoupling from the evolution history of the Universe in the small-scale CMB we observe.
We then provide a detailed prescription for calculating the small-scale CMB in void models (the first peak and beyond), developing the outline first presented in~\cite{RC}. 
Constraints on void models are presented in Sec.~\ref{sec:result}, which is followed by a discussion on implications for other observations.
Sec.~\ref{sec:conc} concludes.
We dedicate the Appendix to describe the importance of including radiation in the analysis, which involves a full two fluid solution to the field equations.

\section{The small scale CMB - a minimal prescription}
\label{cmbth}
\subsection{The FLRW case}

Before we begin to discuss a void model in which to compute the CMB, let us first describe our approach, and re-consider the FLRW case. In a spherically symmetric universe, an observer at the centre observes the CMB at a uniform area distance (also called angular diameter distance) $d_{A}(z_\dec)$ in all directions, at redshift $z_\dec$. Roughly speaking, the temperature fluctuations observed have two contributions: local contributions from the patch of the last scattering surface observed, which depend purely on the physics of decoupling (plus initial conditions set by the primordial power spectrum), and non-local line of sight effects. Let us examine some useful quantities for describing the CMB in FLRW, trying to separate the local from non-local effects, and, in particular, scale factors from redshift.

The CMB shift parameters 
\bea
l_a&=&\pi \frac{d_A(z_{\dec})}{a_{\dec}r_s(a_{\dec})},\nonumber\\
l_{\eq}&=&\frac{d_A(z_{\dec})}{a_{\dec} k_{\eq}^{-1} },\nonumber\\
R_*&=&\left.\frac{3\rho_b}{4\rho_\gamma}\right|_{\dec}=\frac{3}{4}\frac{\Omega_b}{\Omega_\gamma}a_\dec\, ,\label{shift}
\eea
are sufficient to characterize the key features of the first three peaks of the CMB~\cite{Hu:2001bc,Hu2008} (see also~\cite{Wang:2007mza,Komatsu:2008hk,wmap7,Vonlanthen:2010cd} who use different but analogous parameters). Given standard thermal history and matter content, the physics which determine the first three peaks also fix the details of the damping tail~\cite{Hu2008}. With the exception of $d_A(z_{\dec})$, all quantities are local to the last scattering surface of the CMB that we observe, and the evolution of the universe before decoupling. 

Let us consider $l_a$ first. The comoving sound horizon, $r_s(a_\dec)$, is given by
\be
r_s(a_{\dec})=\int_0^{a_{\dec}}\frac{d a}{a^2H(a)\sqrt{3+9a\Omega_b/4\Omega_\gamma}}.
\ee
 Here, the Hubble parameter is given by
\be\label{Fried}
H(a)=H_0\sqrt{\Omega_r a^{-4}+\Omega_m a^{-3}+\Omega_k a^{-2}} ,
\ee
the scale factor $a$ is defined as unity today, and all $\Omega$'s represent today's values; we also define $H_0=100\, h\,\hu$. The full radiation contribution is given by 
\be\label{Omr}
\Omega_r=\left[1+\frac{7}{8}\left(\frac{4}{11}\right)^{4/3}N_{\text{eff}}\right]\Omega_\gamma\,,
\ee 
where $N_\eff$ is the effective number of relativistic degrees of freedom around today~-- the standard model has $N_\eff=3.04$.

Now, $a_\dec=a_\dec(t_\dec)$ is the scale factor at the time of decoupling, $t_\dec$.
 Hence, the proper size of the sound horizon is $a_\dec r_s(a_\dec)$. This means that $l_a$ is the ratio of area distance to proper distance, which are both well defined distance measures in any spacetime~-- unlike comoving coordinate distance which has no physical meaning in general spacetimes (for example, WMAP~\cite{Komatsu:2008hk,wmap7} use comoving distances rather than physical distances).

Now consider $l_\eq$. The equality scale is given locally by the time of matter-radiation equality, when $a_{\eq}=\Omega_r/\Omega_m$. The mode which enters the Hubble radius at equality has comoving wavenumber
\be
k_{\eq}=a_{\eq}H_{\eq}=\sqrt{2\Omega_mH_0^2 a_{\eq}^{-1}}.
\ee
The proper size of the Hubble radius at equality is just $H_{\eq}^{-1}$, so it's useful to think of $l_{\eq}$ as
\be
l_{\eq}=\frac{T_\dec}{T_{\eq}}\frac{d_A(z_\dec)}{H_{\eq}^{-1}}\,;
\ee
again, a ratio of area distance to proper distance. 

For variables which are local to the last scattering surface (LSS) it is possible to write them in terms only of local quantities. In particular, we can remove all reference to `today' where $a=1$ and the $\Omega$'s are defined (or whichever reference time is chosen). Changing variables from scale factor to temperature using $a=T_0/T$, where $T_0$ is today's temperature, we find that all reference to $H_0$, $T_0$ and the $\Omega$'s, etc., factor out of the quantities above, leaving just the baryon fraction $f_b=\Omega_b/\Omega_m$, the local temperature of decoupling $T_\dec$, and the baryon-photon ratio $\eta$ which defined as the number of baryons per photon,
\bea
\eta&=&\frac{n_b}{n_\gamma}=\frac{\pi^4}{30\zeta(3)}\frac{T}{m_p}\frac{\rho_b}{a\rho_\gamma}\nonumber\\
&\approx&6.154\times10^{-10}\left(\frac{T_0}{2.725\,\text{K}}\right)^{-3}\left(\frac{\Omega_bh^2}{0.02258}\right). 
\eea

Thus, the proper size of the sound horizon at decoupling may be written as 
\bea
a_{\dec}r_s(a_{\dec})=\frac{h}{\sqrt{3}H_0T_{\dec}}\int_{T_{\dec}}^\infty dT\,\, T^{-3/2}\times~~~~~~~~~~~~\nonumber\\
\left[(\varpi_\gamma+\varpi_\nu)T + \varpi_b\frac{\eta}{f_b}\right]^{-1/2}\left(1+\frac{3}{4}\frac{\varpi_b\eta}{\varpi_\gamma T}\right)^{-1/2}\label{rs},
\eea
where $h/H_0\approx2998\,\text{Mpc}$ and we have defined the dimension-full constants 
\bea\label{varpis}
\varpi_\gamma&=&\frac{\Omega_\gamma h^2}{T_0^4}
\approx\left(\frac{0.02587}{1\,\text{K}}\right)^4,\\
\varpi_\nu&=&\frac{\Omega_\nu h^2}{T_0^4}\approx0.227 N_\eff\left(\frac{0.02587}{1\,\text{K}}\right)^4,\\
\varpi_b&=&\frac{\Omega_b h^2}{\eta T_0^3}=\frac{30\zeta(3)}{\pi^4}m_p\varpi_\gamma \,. \label{varpi_b}
\eea
Crucially, these have \emph{no dependence on any parameters of the model}. That is, we are not free to specify the $\varpi$'s, apart from $N_\eff$. These are derived assuming that $f_b$ and $\eta$ are constant.\footnote{A similar expression was given in~\cite{Vonlanthen:2010cd} using redshift instead of temperature. Although we don't have dimensionless constants doing it our way, temperature along the worldline up to last scattering is well defined by the local radiation density; redshift is ambiguous in an inhomogeneous universe as it applies along null cones, not timelike worldlines. Furthermore, redshift requires a fixed observation point which we shall find useful to remove where we can.}
Hence, the proper size of the last scattering sound horizon, $a_{\dec}r_s(a_{\dec})$, depends only on $\eta, f_b$ and $T_{\dec}$. The temperature of decoupling at the level of approximation we use also only depends on $\eta$ and $f_b$: The elastic Thompson scattering rate depends only on $\eta$,
\be\label{thompson}
\Gamma(T)=\frac{\sqrt{2\zeta(3)}\sigma_T}{\pi}\eta^{1/2}\left(\frac{m_e}{2\pi T}\right)^{3/4}T^3 e^{-\Delta/2T},
\ee
which at decoupling must equal the Hubble rate
\be\label{H(T)}
\frac{H(T)}{100\,\hu}=\sqrt{(\varpi_\gamma+\varpi_\nu)T^4+\varpi_b\frac{\eta}{f_b} T^3},
\ee
which also only has dependence on the local parameters $\eta$ and $f_b$. (Using the approximation that $\Gamma(T_\dec)=H(T_\dec)$  overestimates the temperature by $2-3\%$~-- we use the Saha approximation for illustration only; see below for the more accurate method we actually use.)
For $l_{\eq}$, the scale which is compared with the area distance to the CMB is
\be
\frac{k_{\eq}^{-1} a_{\dec}}{2998\,\text{Mpc}}=\frac{\sqrt{\varpi_\gamma+\varpi_\nu}}{\sqrt{2}\varpi_b}\frac{f_b}{\eta T_{\dec}}.
\ee
Finally,
\be
R_*=\frac{\varpi_b}{\varpi_\gamma}\frac{\eta}{T_{\dec}}.
\ee

Thus, apart from $d_A(z_\dec)$, the key features of the CMB power spectrum constrain directly $f_b$, $\eta$ and $T_\dec$ \emph{along the radial worldline of the patch of the LSS we observe}. Given an observation temperature of the CMB, these translate into $\Omega_{m,b,r}h^2$ using Eqs.~(\ref{varpis}) --~(\ref{varpi_b}). We have taken the time to make this explicit because the observation temperature is not obvious in void models: that is, we shall find that the temperature entering in those equations is not the temperature observed at the centre of the void, but rather the temperature that a fictitious observer far outside the void would measure.

Instead of the Saha approximation, we will use the fitting formula of~\cite{HS} as a more accurate means to calculate $T_\dec$, adjusted for our purposes. This is normally given as a formula for $z_\dec$ in terms of $\Omega_bh^2$ and $\Omega_ch^2$, and derived for $T_0=2.725\,$K. Using Eq.~(\ref{varpi_b}) we can recast it in terms of $\eta$ and $f_b$:
\bea
T_\dec&\approx&  2855.8\,\text{K}\times \left(  1.0+ 0.078\,{\eta_{10}}^{- 0.738}
 \right) \nonumber\\
&\times& \left[  1.0+{g_1}\, \left(  0.00365\,{\frac {\eta_{10}}{{f_b}}} \right) ^{{g_2}} \right] \label{EH}
\eea
where
\bea
g_1&=& 0.298\,{\frac {{\eta_{10}}^{- 0.238}}{ 1.0+ 0.546\,{\eta_{10}}^{ 0.763}}}\nonumber\\
g_2&=&0.560\, \left(  1.0+ 0.000818\,{{\eta_{10}}}^{ 1.81} \right) ^
{-1}\,,
\eea
and we have defined $\eta_{10}=10^{10}\eta$.

The parameters $l_a,l_\eq,R_*$ capture all the information we need from the small angle CMB (as already mentioned, we are not in a position to evaluate the large-scale features, which includes the late time Integrated Sachs Wolfe effect and so full calculation of the growth of structure). Other `orthogonal' parameter sets are sometimes used. For example, WMAP and others~\cite{Wang:2007mza,Komatsu:2008hk,wmap7,Vonlanthen:2010cd} use $l_a, R$ and $z_\dec$, where 
\bea
R&=&\sqrt{\Omega_mH_0^2}d_A(z_*)/a_*\\
&=&\left(\frac{2\Omega_m}{\Omega_r}\right)^{1/2}l_\eq\\
&=&\left(\frac{2f_bT_0}{\eta}\right)^{1/2}\left(\frac{\varpi_\gamma+\varpi_\nu}{\varpi_b}\right)^{1/2}l_\eq\,.  
\eea
So, this relies on $T_0$, which $l_\eq$ does not. For our purposes this is not good: we shall find that $T_0$ varies considerably on a surface of constant time in void models, which would make this shift parameter awkward to use. Also, the other parameter in this set, $z_*$, is not as neat for us as $T_*$ because it requires a point of observation, rather than just local physics.

\subsection{Inhomogeneous spherically symmetric universe}

As we observe from the above equations, the local part of the CMB constrains only local variables which are meaningful at the time up to decoupling: $\eta, f_b$, and $T_*$ which is fixed by $\eta$ and $f_b$ (or, alternatively, one can obviously choose any orthogonal combination of such variables). Provided the FLRW approximation is valid before decoupling in a horizon size then the equations above which involve $\varpi$'s rather than $\Omega$'s are sufficient. There is no reference to late times in the equations and it does not matter if the spacetime is FLRW after decoupling. 

\begin{figure}[t]
\begin{center}
\includegraphics[width=\columnwidth]{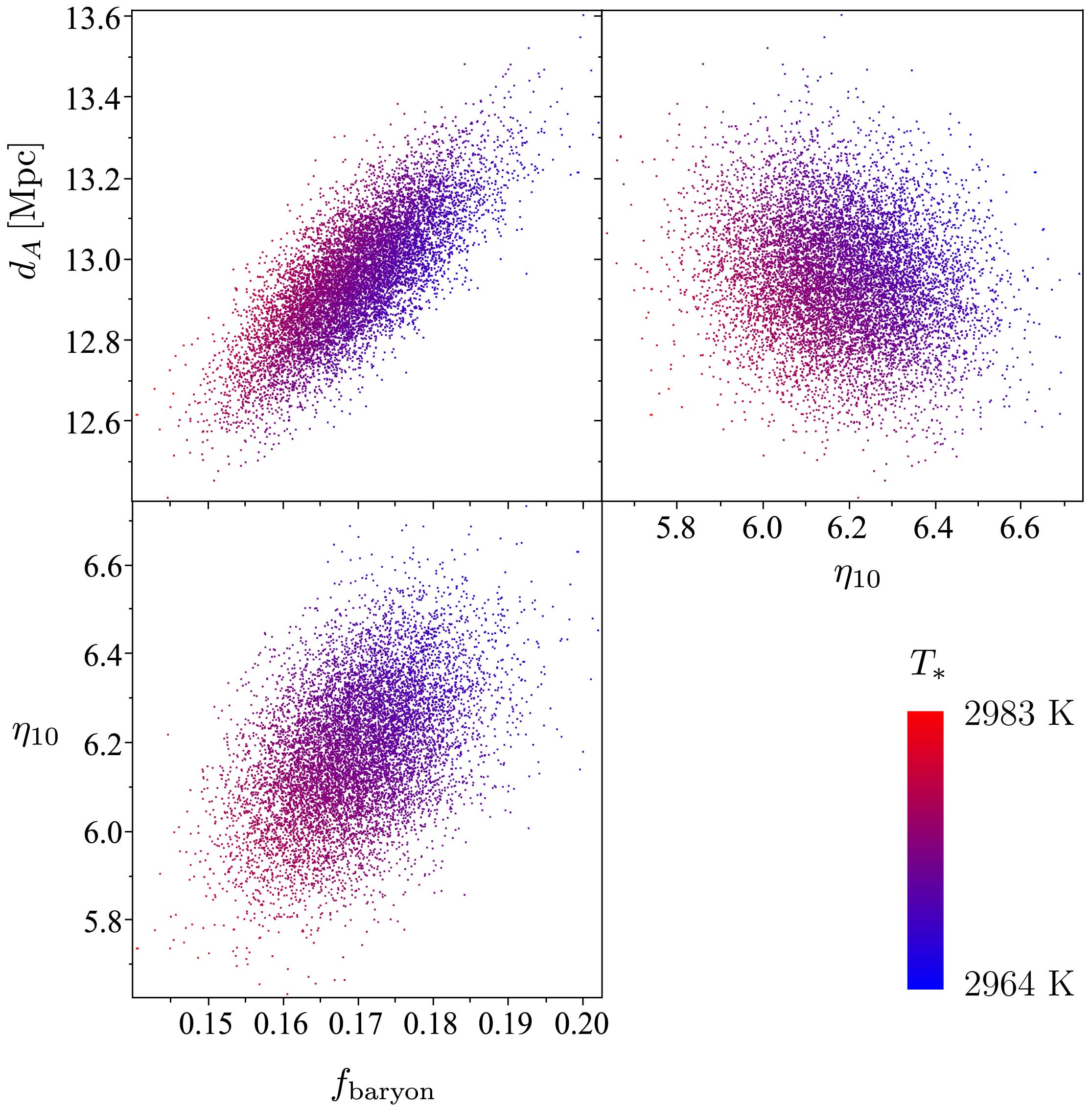}
\caption{Constraints on $f_b$, $\eta$ and $d_A(z_\dec)$ derived from WMAP7 constraints on a $\Lambda$CDM model. The points are sampled assuming Gaussian errors on $l_a=302.44\pm0.8$, $l_\eq=137.5\pm4.3$, $100\Omega_bh^2=2.258\pm0.057$, $z_\dec=1090.79\pm0.92$, the last two of which are used to calculate $R_\dec$. The variables are computed using the fitting formula for $T_\dec$,~Eq.~(\ref{EH}), indicated by the colour. Note the slight degeneracy between $f_b$ and $d_A(z_\dec)$, while the other combinations are essentially independent.}
\label{wmap-fig}
\end{center}
\end{figure}

Let us fix $f_b$ and $\eta$ at the radius of decoupling such that $R_\dec$ and $l_a/l_\eq$ satisfy observational constraints. This implies $T_\dec$ from $R_\dec$ or better from Eq.~(\ref{EH}). Hence, $H_\dec=H(T_\dec)$ is given unambiguously by Eq.~(\ref{H(T)}). In addition, if an observer measures the CMB to have temperature $T_0^\mathrm{obs}$, then the redshift of the CMB will be $1+z_\dec=T_\dec/T_0^\mathrm{obs}$, since the black-body spectrum is conserved in any space-time~\cite{EMM}. The final constraint from the CMB parameters then must be $d_A(z_\dec)$, which follows from $l_a$ or $l_\eq$. For a spherically symmetric universe observed from the centre, $d_A(z_\dec)$ is just one constraint on the model. The current WMAP7 constraints on the three degrees of freedom are illustrated in Fig.~\ref{wmap-fig}.
A slight degeneracy is present between $f_b$ and $d_A$. Observational data require $\eta_{10}\approx6.2$, $f_b\approx0.17$, and $d_A\approx13$ Mpc (which will be used as benchmark values in some of the following figures).

\begin{figure*}[t]
\begin{center}
\includegraphics[width=0.9\textwidth]{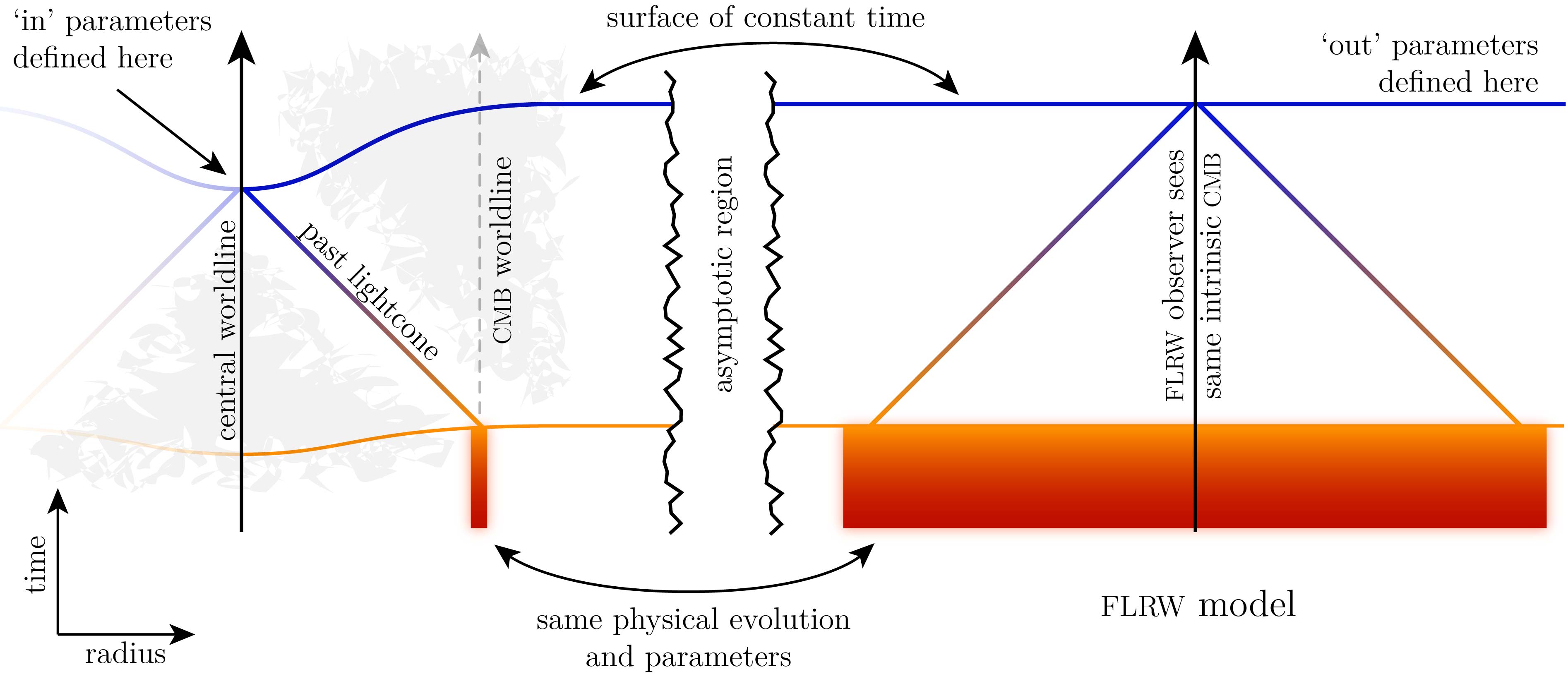}
\caption{Schematic of our model and assumptions. The central observer sees a small patch of the last scattering surface, which we assume evolves as in the asymptotic part of the model which is FLRW. The shaded regions are not required to be fixed for the calculation of the small scale CMB, but are constrained a posteriori.}
\label{overview}
\end{center}
\end{figure*}

We now want to link such constraints to the parameters of the model.
For our purposes, we can calculate nearly everything along the central worldline and along one at the radius of the CMB ($r\gg$ void width), where we assume an FLRW description holds with separate parameter values at the centre and asymptotically, denoted $\in$ and $\out$, illustrated in Fig.~\ref{overview}.
The exception is the area distance to the CMB, as discussed below. 

The parameter values labelled `(out)' are, strictly speaking, asymptotic values and are not necessarily the values of the parameters along the worldline where we observe the LSS (which we refer to as the `CMB worldline' located at coordinate distance $r_\dec$). Since we consider inhomogeneities which are Gpc in size, and the LSS is located where the profile is flat~\footnote{There may be effects from the slight radial gradient at decoupling. The profile can be chosen exactly flat at $r_\dec$ if required. As we are neglecting the bang time function, any other type of inhomogeneity should also be small at this time.}, decoupling occurs in the asymptotic FLRW region. However, an observer at that radius will eventually see the inhomogeneity, and so the FLRW approximation will no longer be valid.

Our approach relies in some respects on the spacetime at the centre being locally (very close to) FLRW, but we do not rely on this for our key results. We give a discussion of this in the appendix.

\subsection{The CMB calculation for a void}\label{cmbcalc}

Voids may be described by a profile for the present-day matter density parameter. We shall typically consider a Gaussian as a benchamrk case (although the void shape is only mildly constrained by SNIa at low redshift, e.g.~\cite{FLSC}, and we discuss its impact on obtained result in Sec.~\ref{sec:da}) : 
\bea
\Omega_m(r)&=&\omout-(\omout-\omin)\,e^{-r^2/(2\,\sigma^2)}\nonumber\\
&=&1-\Omega_k(r)-\Omega_r(r),
\eea
and we can assume a similar profile for the radiation density $\Omega_r(r)$ (usually ignored but important here). 
The Hubble rate is given by a generalized Friedmann equation, which we discuss in Appendix~\ref{sec:rad}. When in the matter era and radiation can be safely ignored this is given by the usual LTB relation (see, e.g.,~\cite{enqvist}).

Along the central worldline and along one at the radius of the CMB ($r\gg\sigma$), we can neglect all $r$-dependence, and use the standard Friedmann equation, Eq.~(\ref{Fried}) with separate $\in$ and $\out$ parameter values.
The area distance to the CMB is instead calculated in the normal way for an LTB model~\cite{enqvist,FLSC}, and is affected by the shape and width of the void.
The CMB places constraints on the void model through Eqs.~(\ref{varpis})~-- (\ref{varpi_b}), which hold provided the model is asymptotically FLRW, as well as the area distance to the LSS. The remaining part of the model is specified by local observations of the CMB temperature, the Hubble rate, SNIa observations and so on.

To fully fix the model we need to specify the parameters
\be
T_0,~h,~\Omega_{m},~f_b,~\eta,~N_\eff,
\ee 
at the centre and asymptotically. We choose $a(t_0,r)=1$ to fix our gauge.  
We constrain the void as follows:\\

\noindent{\bf Inside: the centre of the void}

Core constraints:
\begin{itemize}

\item From the CMB temperature today we have $$T_0^\in\approx2.725K\Rightarrow \Omega_\gamma^\in h_\in^2\approx 2.469\times10^{-5}.$$ This gives $\Omega_r^\in\approx1.69\Omega_\gamma^\in$ for $N_\eff=3.04$ from Eq.~(\ref{Omr}).

\item The local expansion rate should be taken from local observations only, unless a full likelihood analysis is performed. For example,~\cite{Riess:2009pu} find $h^\in=0.742\pm0.036$ from low redshift SNIa.  

\item A void model which fits the SNIa well can have very low density at the centre.  We can expect $$\omin\sim 0.1-0.2$$ from SNIa constraints. \footnote{Age constraints favour a slightly lower value of $h$ than the \cite{Riess:2009pu} value, but with $\omin$ this low, recent constraints~\cite{ages} are relatively easy to satisfy. We don't consider this further here.}
\end{itemize}

Additional parameters:

\begin{itemize}

\item The baryon-photon ratio can vary radially, reflecting the initial radiation distribution. From $^7$Li measurements we can set (for models such that $\eta$ is conserved or nearly conserved)~\cite{RC} $$\eta^\in\sim4-5\times 10^{-10};$$ alternatively, we can fix $\eta$ as a constant and determine it directly from the CMB. Note that $\eta^\in$ refers to the baryon-photon ratio at late times at the centre, which is not the primordial value if $\eta$ is not conserved (see the Appendix).

\item From Eq.~(\ref{varpi_b}) we can find
\be\label{obin}
\Omega_b^\in = \varpi_b \frac{\eta^\in {T_0^\in}^3}{h_\in^2}.
\ee
This fixes the local baryon fraction at the centre $f_b^\in=\Omega_b^\in/\omin$. An obvious choice is to make the assumption that $f_b=$const. so that $f_b^\in=f_b^\out$. Given that the CMB constrains $f_b^\out$, this puts an additional constraint on $\Omega_m^\in$. Local observations of $f_b$ could instead be folded in at this stage; for example, from observations of gas in clusters one finds $f_b\gtrsim0.11$~\cite{Allen:2007ue}, with a spatial variation $\lesssim 8\%$ up to $z\sim1$~\cite{Holder:2009gd}.

\end{itemize}

These last two are not necessary for the CMB analysis, but are useful to define nonetheless.

Now to make the link with the asymptotic parameter values, we
calculate $$t_0^\in=\int_0^1\frac{da}{aH}$$ along the central worldline. We assume the Bang time to be homogeneous $t_B=0$, so $t_0^\in=t_0^\out$ is the age of the universe everywhere today (the bang time function is a decaying mode in LTB models~\cite{silk}). This fixes the time, at $a(t_0)=1$, where the asymptotic values of quantities are defined.

Note that the only parts of the calculation where we have used the assumption  that the spacetime along the central worldline is FLRW lies in calculating the baryon fraction on the inside, $f_b^\in$, and the overall age of the model. Thus, in cases where this is not satisfied, this will require an adjustment of $f_b^\in$ from the values we give below.
\\

\noindent{\bf Outside: asymptotic parameter values}

\begin{itemize}

\item Given values of $l_a/l_\eq$ and $R_*$ we find values for $\eta^\out$ and $f_b^\out$, and, hence,  $T_\dec^\out$ from Eq.~(\ref{EH}). 

\item The redshift of the LSS is 
\be
1+z_\dec=\frac{T_\dec^\out}{T_0^\in}\;.
\label{eq:redT}
\ee

\item If we ignore the contribution from radiation, the age of the universe, $t_0^\out$, fixes $\Omega_m^\out$ given $h^\out$ or $h^\out$ given $\omout$. For arguments sake, let's assume we choose $\Omega_m^\out$ and fix $h^\out$ from this. (The radiation density makes virtually no difference to the age.)

\item The observed area distance to the LSS, $d_A^\mathrm{obs}(z_\dec)$, derived from $l_a$ or $l_\eq$, may be approximately calculated in an LTB void model, neglecting the effect of the radiation (we discuss the errors from this approximation in the Appendix). This constrains the void profile as well as $\Omega_m^\out$ (given $h^\out$ from $t_0$) but fixes neither uniquely. 

\item Correlation with SNIa data may be used to break the degeneracy on $\sigma$ and $\Omega_m^\out$. (SNIa favour $\text{FWHM}\sim6\,$Gpc and $\Omega_m(z\sim1)\sim0.5$ when $\omin\sim0.1$~\cite{FLSC}.)

\item We now calculate $T_0^\out$ from Eq.~(\ref{varpi_b}). We can now find $\Omega_{\gamma,r}^\out$ trivially (given $N_\eff$). Recalculating the age can now refine the estimate of $h^\out$. (We find these precisely iteratively.)

\end{itemize}

A final constraint arises from the measurement of $T_0^\in$ and its link to $T_\dec^\out$ through Eq.~(\ref{eq:redT}), namely when we integrate out along the past lightcone from the centre out to $z_\dec$. In terms of time rather than temperature, it says that the local time at that point must equal the time obtained by integrating up along the timelike worldline from the big bang up to decoupling. That is,
\be
t_0-t_\dec(z_\dec)=\int_0^{z_{\dec}}\frac{d z}{1+z}\left.\left({\partial_t\ln\chi}\right)^{-1}\right|_{\text{nullcone}}\,,\label{glue}
\ee
where $\chi(t,r)=-d t/d r$ evaluated on the past nullcone, and $t_\dec(z_\dec)$ is the local time of decoupling at the redshift observed from the centre, which must be equal to 
\be
t_\dec=\int_{T_\dec}^\infty \frac{dT}{T}\frac{1}{H(T)}\,,
\ee  
where $H(T)$ is given locally by Eq.~(\ref{H(T)}) (with no reference to late times).
Because the integral in Eq.~(\ref{glue}) traverses the void, it is a constraint on the profiles. This constraint is extremely sensitive to the details of the radiation inhomogeneity (and the `out' parameter values) because most of the integral cancels the $t_0$ on the l.h.s..\footnote{Indeed, if we evaluate this constraint in FLRW, $t_*$ varies by tens of percent depending on the amount of radiation present.} Indeed, since $t_\dec/t_0\sim10^{-5}$, we need to know the whole spacetime to better precision than this; matching an LTB void to a radiation filled FLRW model cannot achieve this accuracy (see the Appendix).  If the central worldline is FLRW, then this constraint must be automatically satisfied and can be inverted to find the density profile of the radiation between $\in$ and $\out$. For the models considered here, this typically gives an inhomogeneity $\mathcal{O}(1)$ in the radiation density at early times (see Fig. 2 of~\cite{RC}). We expect that as the void size decreases this will lead to stringent constraints on the radiation inhomogeneity; but for the Gpc scale voids we are interested in we expect the constraints to be mild since the void scale is only entering the Hubble scale today. Essentially, this will tell us how to map the initial radiation profile onto the central worldline evolution for the temperature.  
 
However, Eq.~(\ref{glue}) deserves a more detailed discussion, which is presented in Appendix~\ref{sec:rad}. 
Indeed, one of the main differences between this work and previous analyses resides in the introduction of inhomogeneous radiation which can significantly affect this $t_\dec(z_\dec)$ relation (or, equivalently, $a_\dec(z_\dec)$) with respect to the relation computed within the LTB framework matched to a radiation filled FLRW model. In essence, $z_*$ is fixed by the CMB, but $t_*$ and $a_*$ depend on whether radiation is included or not. This changes things considerably. Indeed, considering Eq.~(\ref{glue}) as a constraint on $t_0$ \emph{given} $z_\dec$ and $t_\dec$, this effect has similar consequences to the introduction of an inhomogeneous bang-time function since $t_\dec$ is the time since bang in that context.

All the key features of the void are now known and we can derive quantitative constraints.

\section{Results}
\label{sec:result}

\subsection{Constraints from physics of decoupling}
\begin{figure*}[ht]
\begin{center}
\includegraphics[width=1.0\textwidth]{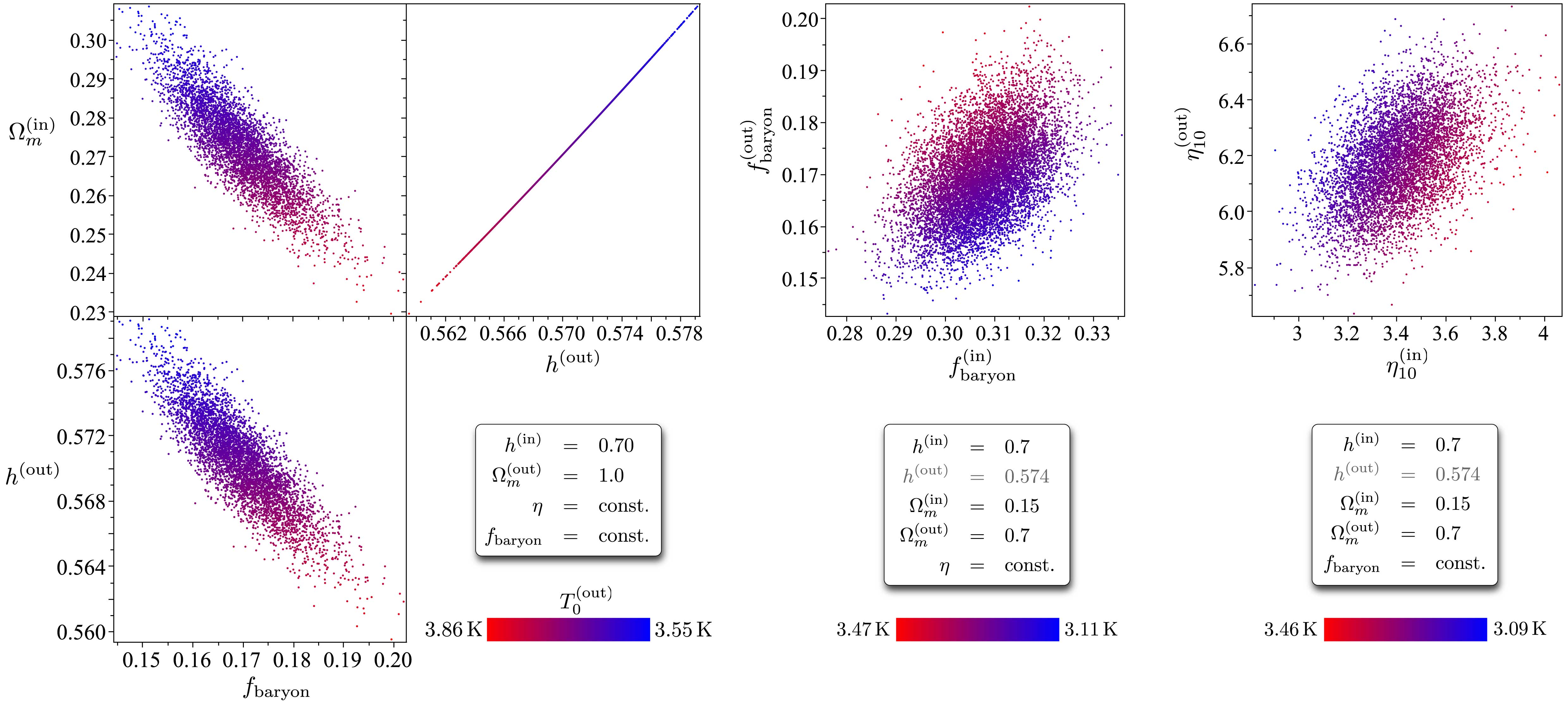}
\caption{Constraints on an asymptotically flat void, ignoring the area distance to the CMB (left). Models which fit $l_a/l_\eq$ and $R_\dec$ are picked as in Fig.~(\ref{wmap-fig}), which constrains $f_b$ and $\eta$, which we assume spatially constant in this case. Because $\eta$ is constant rather than fixed at the centre, this results in a scatter in the void parameters shown.  In the middle, we have chosen an asymptotically open void, which lowers $T_0^\out$, and we have released the assumption on the baryon fraction; instead we have chosen the matter density inside as fixed. The baryon fraction at the centre now takes up the scatter in values from the CMB, resulting in a fixed void with fixed $h^\out$ (derived). Similarly, we can choose the baryon fraction fixed and let $\eta^\in$ take up the scatter~-- shown right~-- in which case these parameters favour a low value of $\eta^\in$, in line with the lithium constraints~\cite{RC}.}
\label{void-fig1}
\end{center}
\end{figure*}

First, we consider constraints arising the from physics of decoupling, namely from $l_a/l_\eq$ and $R_\dec$ parameters. Here we do \emph{not} take into account the bounds from the area distance.
As an illustration of how the CMB directly constrains the model, consider the left panel of Fig.~\ref{void-fig1} where we take an asymptotically flat void with $h^\in=0.7$. Assuming both $\eta$ and $f_b$ constant, we see that the CMB constraints translate into constraints on $\omin$ and $h^\out$, as well as $T_0^\out$. For this kind of void, $\omin$ turns out to be quite large, and not easily compatible with SNIa data (moreover, the area distance is too low unless the void profile has a dip of very low density in it, which would be again tough to fit to SNIa; see below).  
Releasing the assumption of a constant baryon fraction or constant baryon-to-photon ratio, one can fix the matter density inside. In the first case (middle panel, now in a model which is open asymptotically), the derived $f_b^\in$ is significantly larger than $f_b^\out$ and in possible tension with cluster data~\cite{Holder:2009gd}. When $\eta$ is allowed to vary instead (right panel), $\eta_{10}^\in\approx3.5$ fits the CMB, which may also give an explanation to the $^7$Li problem of standard cosmology as proposed in~\cite{RC} (actually, $^7$Li constraints favour a slightly higher value than this).

\begin{figure}[ht]
\begin{center}
\includegraphics[width=0.8\columnwidth]{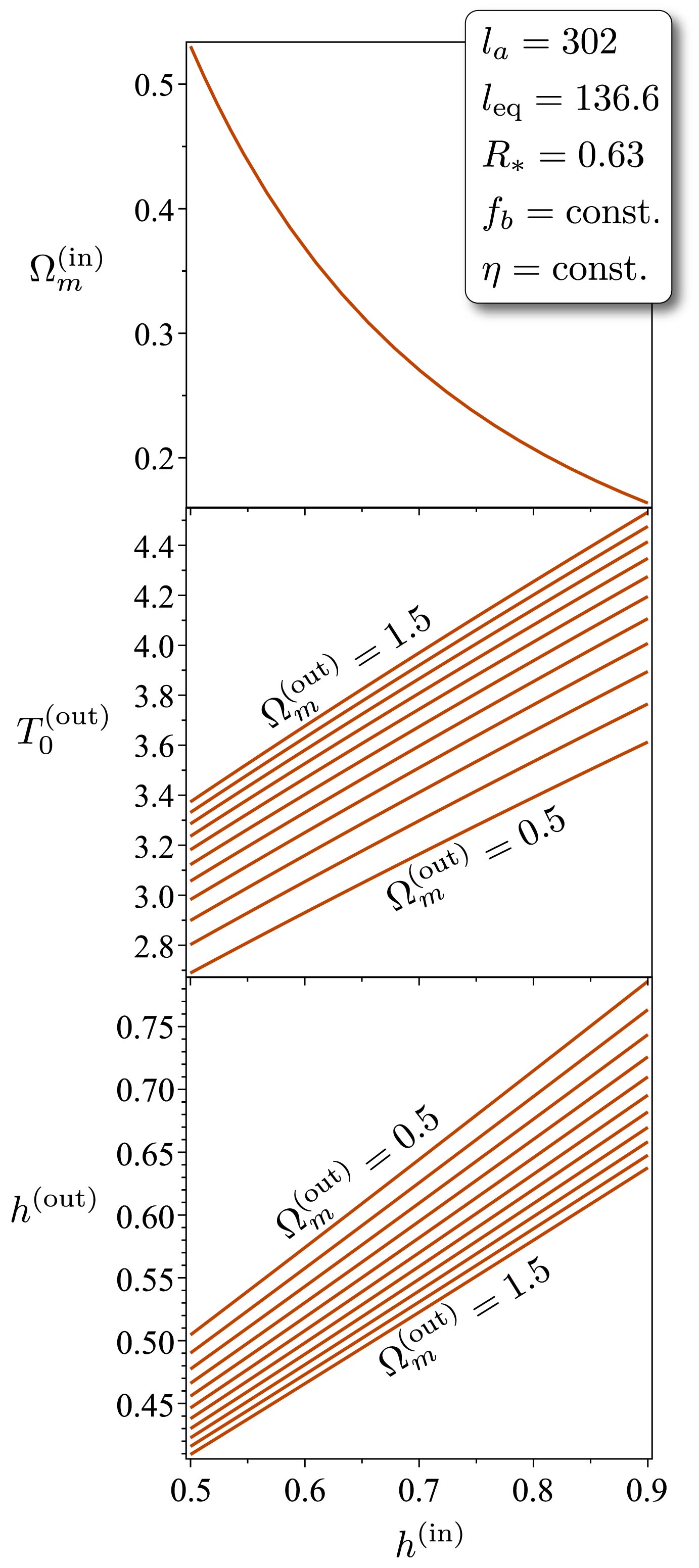}
\caption{Constraints on a general void as a function of the Hubble rate at the centre, ignoring the area distance to the CMB. Here we have chosen the CMB parameters as indicated which imply $f_b=0.17$ and $\eta_{10}=6.2$, and we have chosen these variables as constant. Different choices of $\omout$ then give different $h^\out$ and $T_0^\out$, but $\omin$ (top) is constrained purely by the assumption on the baryon fraction. Note that models with $T_0^\out\simeq T_0^\in$ have \emph{very} low $h^\in$.}
\label{void-fig2}
\end{center}
\end{figure}
\begin{figure}[ht]
\begin{center}
\includegraphics[width=0.8\columnwidth]{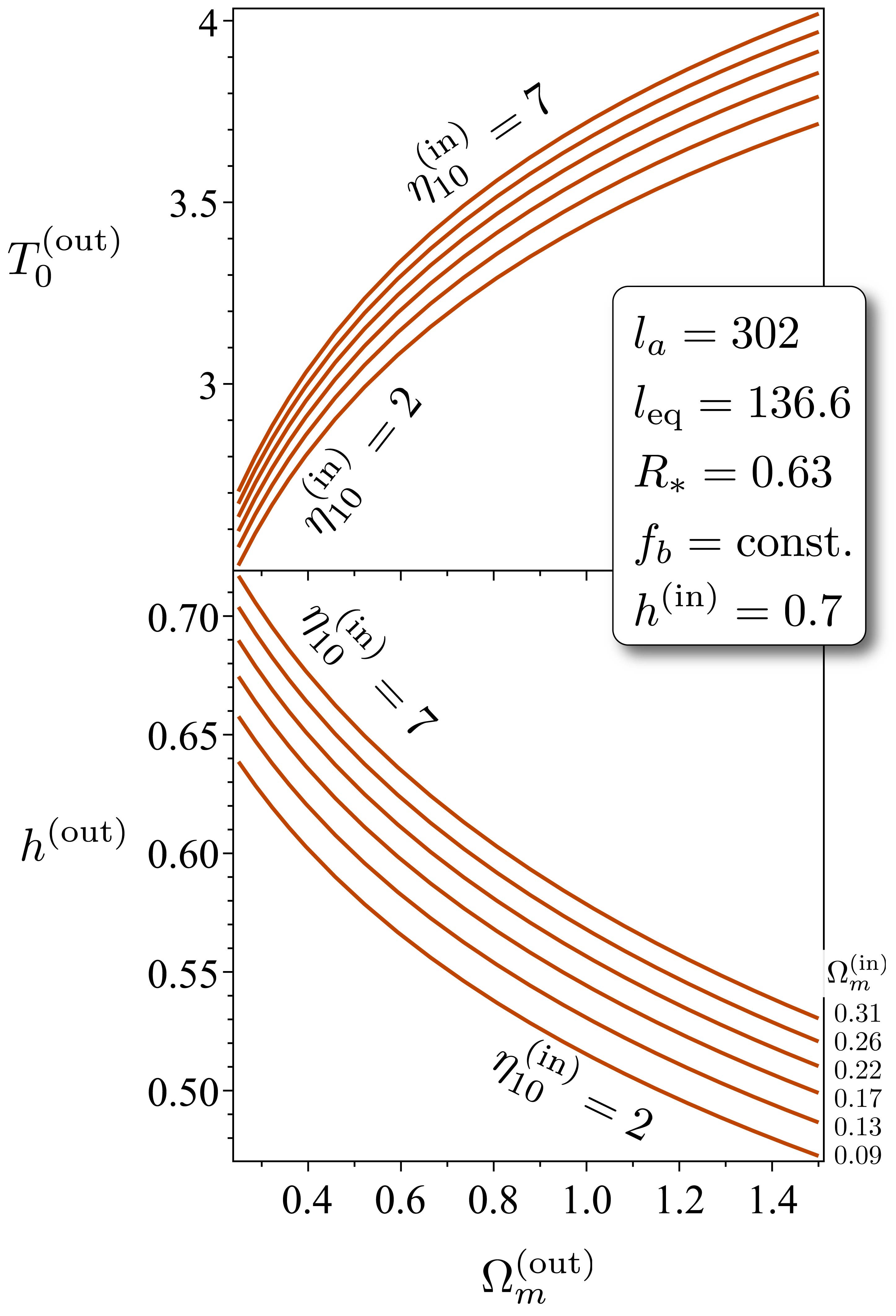}
\caption{Constraints on a general void as a function of asymptotic matter density, ignoring the area distance to the CMB. Here we have chosen the CMB parameters as indicated which imply $f_b=0.17$ and $\eta_{10}^\out=6.2$. We have allowed $\eta^\in$ to take on different values, which, when the baryon fraction is constant, implies differing $\Omega_m^\in$ (indicated).}
\label{void-fig3}
\end{center}
\end{figure}

We gain a more systematic understanding of the constraints implied while ignoring the area distance in Figs.~\ref{void-fig2}~-- \ref{void-fig3}. We choose some CMB parameters $l_a/l_\eq, R_*$, which imply $f_b^\out$ and $\eta^\out$, and show how these constraints translate into the parameters of void models. 

Consider the case when $f_b$ and $\eta$ are both constant, as in Fig.~\ref{void-fig2}. As we vary $h^\in$, $\omin$ is fixed by the assumption on $f_b$ through Eq.~(\ref{obin}) (top panel). 
Other parameters such as $T_0^\out$ and $h^\out$ depend on $\omout$ and $h^\in$. 
Indeed, $h^\out$ is set by the age $t_0^\out$ (assuming a homogeneous bang time). The bottom panel shows that larger $h^\in$ means shorter age and so larger $h^\out$, and that larger curvature $\Omega_k^\out$ (i.e., lower $\omout$) would increase the age and so needs to be compensated by a larger $h^\out$.
The $T_0^\out$ dependencies (central panel) follow again from Eq.~(\ref{obin}), namely, radiation density scales linearly with matter density when $f_b$ and $\eta$ are constant.

In Fig.~\ref{void-fig3} we release the assumption that $\eta=$constant.
Increasing $\eta^\in$ implies, for a fixed $h^\in$ and $T_0^\in$, a larger $\omin$, which leads to a shorter age, and, in turn, for a given $\omout$, to higher $h^\out$ (lower panel) and $T_0^\out$ (top panel, see Eq.~(\ref{obin}) for fixed $f_b^\out$ and $\eta^\out$). The $h^\out$ vs. $\omout$ relation is given by the age constraint as in Fig.~\ref{void-fig2}, and $T_0^\out$ vs. $\omout$ is again related to Eq.~(\ref{obin}).

\subsection{Constraints including the area distance}\label{sec:da}

\begin{figure}[ht]
\begin{center}
\includegraphics[width=0.95\columnwidth]{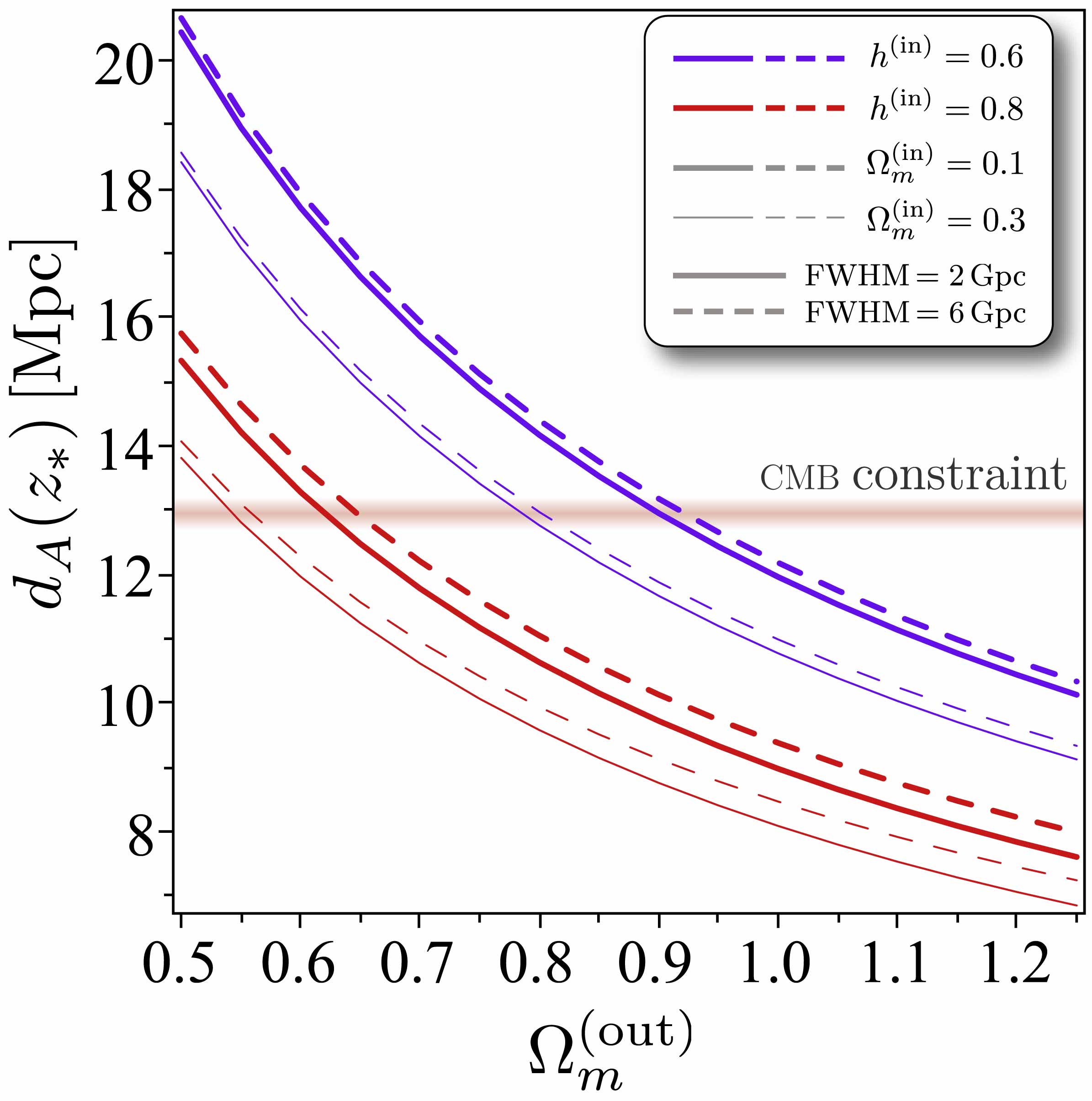}
\caption{Area distance to the CMB as a function of asymptotic matter density. Here we have used CMB parameters such that $z_\dec=1090$. Clearly, there is no problem making a void model with the correct area distance to the CMB, even with a high value of $h^\in$, provided the model is open at the CMB radius. The CMB constraint is shown for $N_\eff=3.04$; increasing this lowers the required area distance to the CMB.}
\label{void-fig4}
\end{center}
\end{figure}

Now let us consider the area distance to the CMB. This mainly depends on the void profile and the Hubble rate at the centre (and $z_\dec$). In Fig.~\ref{void-fig4} we see how, changing $h^\in$, $\omin$, $\omout$ and the width $\sigma$. In essence, the lower the density, the higher the curvature, and the larger the area distance to a fixed redshift. The longer the null geodesic spends in a low density region, then, the further the CMB will be located. The CMB constraint of $d_A(z_\dec)\simeq13\,$Mpc implies (with the caveat that we are assuming $d_A$ not to be affected by radiation inhomogeneity) that for a high Hubble rate at the centre the matter density asymptotically must be less than unity~-- an asymptotically flat void is very difficult to achieve, in agreement with~\cite{ZMS,CFZ}~-- and the width must be several Gpc (also implied by SNIa). Lowering the central Hubble rate pushes up the curves in Fig.~\ref{void-fig4}. For a very low central Hubble rate a flat or closed model asymptotically is required to fit $d_A(z_\dec)$.

To summarize, then, let us consider how to use these figures to estimate a specific case. Let's choose $h^\in=0.7$, and $\eta^\in_{10}=4$ to coincide approximately with lithium constraints. Then, if the baryon fraction is constant, we observe from the lower panel of Fig.~\ref{void-fig3} that we must be on the curve with $\omin\approx0.17$ (which is fine for SNIa). Turning to Fig~\ref{void-fig4}, we can see that for this matter density, and, say, a FWHM$\,\sim2$Gpc, we must have $\omout\sim0.7$. Returning to Fig.~\ref{void-fig3}, we have $h^\out\approx0.58$ and $T_0^\out\approx3.3$. A void with these parameters will give a decent fit to the CMB. 

In the recent work of Refs.~\cite{BNV,MZS}, it has been found that the CMB constraints are satisfied for asymptotically closed models with $h^\in\lesssim0.5$.
It's not difficult to extrapolate this case from Fig.~\ref{void-fig4} and to note that we also find the area distance is correct.
Although such models are not specifically included in Figs.~\ref{void-fig2}~-- \ref{void-fig3}, they are constructed such that $f_b^\out\approx0.17$ and $\eta^\out_{10}\approx6.2$, which implies, given what we said above, that they also satisfy constraints from physics of decoupling.
Therefore, following our approach, these models constitute a subclass (i.e., with nearly homogeneous radiation, see Appendix) among the void models which can fit the CMB (and we do not particularly emphasize them mainly because of the low $h^\in$ which is tension with data~\cite{Riess:2009pu}).

\subsubsection*{An asymptotically flat void?}
\label{secflat}

A key finding of previous work is that the CMB rules out asymptotically flat voids~\cite{ZMS,CFZ}, unless the bang time is varying, or the matter density or Hubble rate are much lower than otherwise observed. Under our approximations for the computation of $d_A$, we also find the same result, principally because an asymptotically flat model with sensible $h^\in$ and $\Omega_m^\in$ has the CMB far too close to us, if the CMB worldline is in the asymptotic region~-- this can easily be inferred from Fig.~\ref{void-fig4}. Is there any way to make an asymptotically flat model fit?

It has been shown in \cite{BW,Yoo:2010qy} that one can add features to the radial profile to alter the area distance to the CMB. For example, since SNIa probe the distance modulus only out to $z\sim2$, there are no distance observations between them and the CMB. This freedom can be utilised to put an under-dense shell around us, thereby increasing the distance to the CMB by the required degree. 

There is another way to make asymptotically flat models fit. If we increase the number of relativistic degrees of freedom above the canonical value of $N_\eff=3.04$ (and allow for a spatially varying $f_b$), then constraints in Fig.~\ref{wmap-fig} will be modified and, in particular, this can bring down the area distance to the CMB.
We find that with $N_\eff\sim8$ an asymptotically flat model, with $h^\in\approx0.7, \omin\approx0.2, \text{FWHM}\approx4\,$Gpc fits the CMB perfectly well. This is a similar conclusion to~\cite{huntsarkar}  and deserves further investigation in this context.

\begin{figure}[hb]
\begin{center}
\includegraphics[width=1\columnwidth]{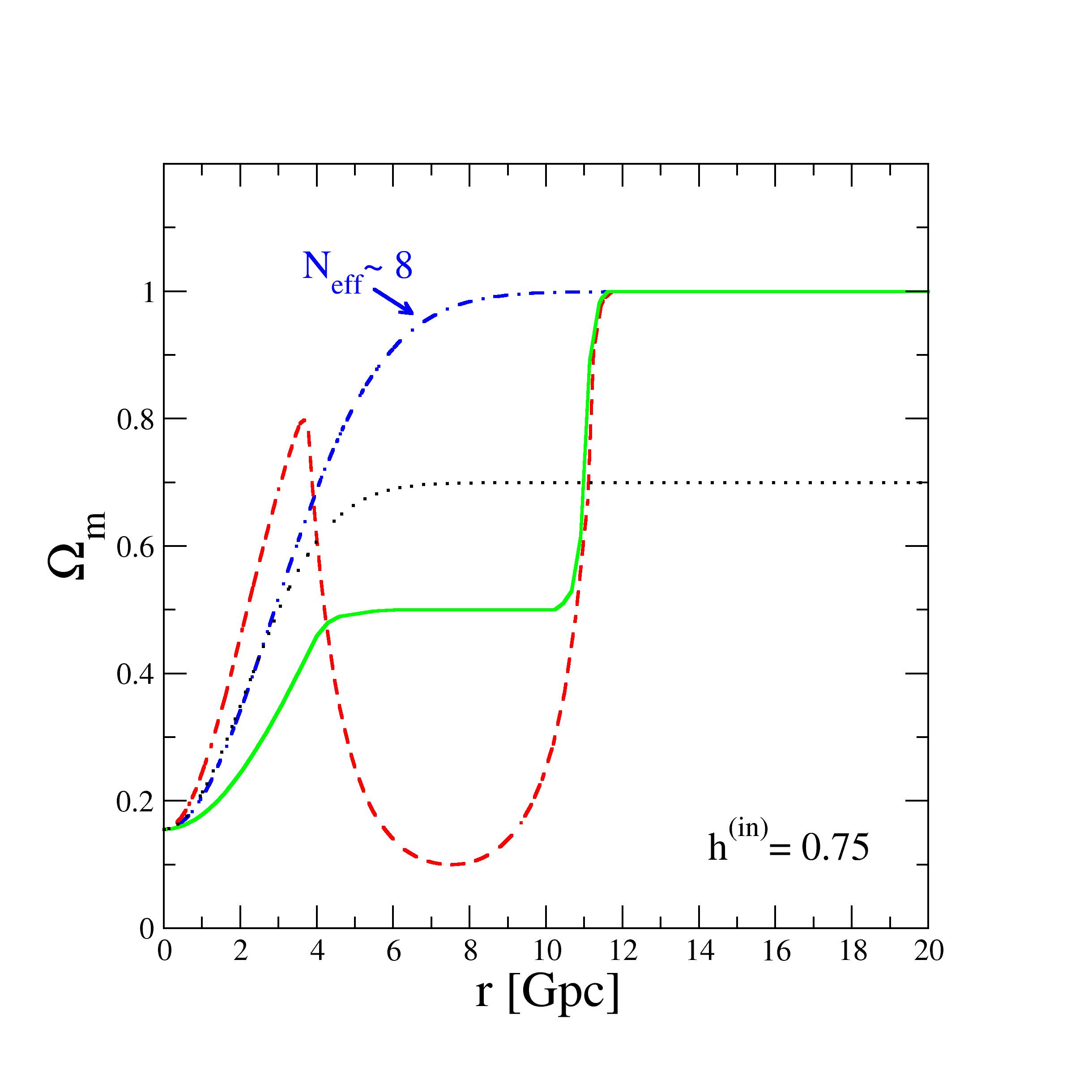}
\caption{Examples of inhomogeneous models with a central void providing an area distance to LSS able to fit CMB data. The computation of $d_A$ has been performed assuming LTB framework. Open models (black dotted) naturally fit. Asymptotically flat models either requires a profile with a shape more complicated than a simple Gaussian void (red dashed and green solid) or an $N_{\eff}$ significantly larger than $3$ (blue dashed-dotted).}
\label{fig:profiles}
\end{center}
\end{figure}

These scenarios are depicted in Fig.~\ref{fig:profiles}.

\subsection{CMB power spectrum}

Once the void model is fully specified the power-spectrum of the CMB may be calculated properly using the procedure given in~\cite{CFZ,Vonlanthen:2010cd}, for large $\ell$. 
This is an important final step because the observational constraints on the 3 parameters we have used to constrain the model, $l_a,l_\eq,R_*$, are actually derived parameters from the data, using an FLRW model~\cite{wmap7}. So, a void model which gives particular values for  $l_a,l_\eq,R_*$ may have a slightly different CMB power spectrum from an FLRW dark energy model with the same  $l_a,l_\eq,R_*$. We find this difference of order a percent or so, in broad agreement with using the parameters for different dark energy models~\cite{Wang:2007mza}. 

To calculate the CMB angular power spectrum for a given void model, first calculate the $C_\ell$'s for an FLRW model using the asymptotic parameter values. This power spectrum will have the same intrinsic CMB features as the void model, since they see the same LSS, but will need to be shifted to account for the incorrect area distance used in the calculation. Let the area distance in the FLRW model be $\hat d_A(\hat z_\dec)$, where the LSS is at redshift $\hat z_\dec$. The relation to convert to the power spectrum the void observer will see is~\cite{Vonlanthen:2010cd} (see also~\cite{CFZ,Zibin:2007mu})
\bea
C_\ell&=&\sum_{\hat\ell}\left(\hat\ell+\frac{1}{2}\right)\hat C_{\hat\ell}\int_0^\pi \sin\theta d\theta\times \nonumber\\
&&P_{\hat\ell}\left[\cos\left(\theta\frac{d_A(z_\dec)}{\hat d_A(\hat z_\dec)}\right)\right]P_\ell(\cos\theta)\nonumber\\
&\simeq& \left[\frac{\hat d_A(\hat z_\dec)}{d_A(z_\dec)}\right]^2
\hat C_{\left[{\hat d_A(\hat z_\dec)}/{d_A(z_\dec)}\right]\ell} \text{~~~for~~~}\ell\gg1\,,
\eea
where it is assumed that $\hat C_\ell$ is smoothed appropriately for non-integer $\ell$. 

As the final step to computing the CMB power spectrum, we can use CAMB~\cite{CAMB} as follows. First compute a void model which is going to give appropriate CMB parameters, according to the steps given above. Then compute the CMB power spectrum using the `out' parameters for $h$, $\Omega_m h^2$, $\Omega_b h^2$ and $T_0$ in CAMB (using a zero $\Lambda$ model). Compute $\hat d_A(\hat z_\dec)$ for this model. The shift parameter $S=d_A(z_\dec)/\hat d_A(\hat z_\dec)$ may be calculated with $d_A(z_\dec)$ found from $l_a$ (since $\sigma$ can be tweaked to make this precise). Then, the output of CAMB, which gives $\ell(\ell+1)C_\ell$, requires the $\ell$ axis compressed by $S$ with no shift to the amplitude, since, using $S=\ell/\hat\ell$, 
\be 
\ell(\ell+1)C_\ell\simeq [\hat \ell(\hat\ell+1)S^2][S^{-2}\hat C_{\ell/S}]\simeq \hat\ell(\hat\ell+1)\hat C_{\hat\ell} \,.
\ee

\begin{figure}[th]
\begin{center}
\includegraphics[width=\columnwidth]{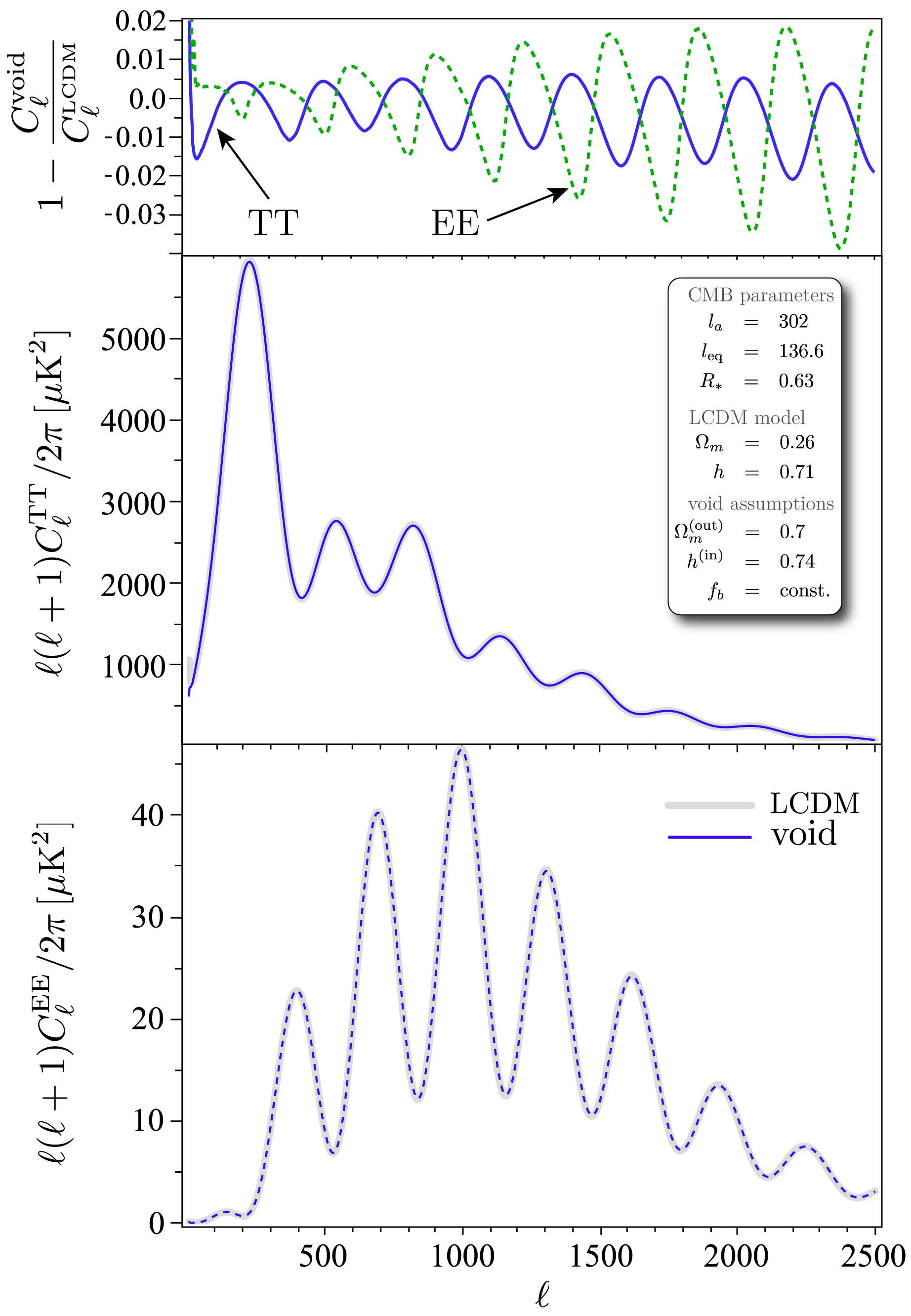}
\caption{The TT (middle) and EE (bottom) angular power spectra for a flat $\Lambda$CDM model and a void model which give the same $l_a,l_\eq$, and $R_*$ parameters. The difference between the two is just a few percent (top). Although the void is derived with the assumptions indicated, we find similar plots for different types of void which fit the CMB parameters indicated.}
\label{cls}
\end{center}
\end{figure}
As an example, let us compare with a flat $\Lambda$CDM model which fits the WMAP7 data well. Choose $l_a=302, l_{\eq}=136.6$ and $R_*=0.63$ which give a good fit. Solve Eqs.~(\ref{shift}) for the flat $\Lambda$CDM parameter values. This gives: $\Omega_bh^2\approx0.0226, \Omega_mh^2\approx0.133$, and $h=0.713$, in good agreement with the WMAP7 best fit. We show the TT and EE power spectra for this model with zero tilt in the primordial spectrum in grey in Fig.~\ref{cls}. Let us compare this with a void model which has the same parameters $l_a$, $l_{\eq}$ and $R_*$. As an example, let us choose a model with constant $f_b$, open asymptotically with $\Omega_m^\out=0.7$, with a high Hubble rate at the centre to fit HST data $h^\in=0.74$, and $\sigma\approx6\,$Gpc. Derived parameters with these assumptions, and taking a constant $\eta$, are $\Omega_m^\in\approx0.242$ (or, alternatively, $\Omega_m^\in\approx0.156$ assuming $\eta^\in_{10}=4$), $\Omega_b^\out\approx0.120$, $h^\out\approx0.636$ and $T_0^\out\approx3.15\,$K, the last 3 of which are used directly in CAMB. Using a shift parameter of $S\approx0.868$, we find the CMB TT and EE power spectra shown in blue in Fig.~\ref{cls}. As we can see there is very little difference between the FLRW model and the void model. Nearly identical figures are found for other void models which produce the same $l_a,l_\eq$, and $R_*$ parameters.

\section{Discussion}
\label{sec:disc}

The CMB in inhomogeneous universe models is clearly complicated (see the Appendix). Even when spherical symmetry is assumed the field equations are too difficult to give a tractable solution when radiation and matter are to be accounted for. The reason is that because the density is not homogeneous, pressure gradients in the radiation force the radiation and matter to have different rest-frames and become non-comoving after decoupling. In fact, CDM decouples from the radiation and baryons much earlier, and so we really have to deal with a three-fluid solution to properly model the spacetime~-- and that is before perturbations are included! As we have discussed, these effects appear to be critical for a complete analysis of the CMB. Indeed, it is important to note that the void amplitude is significantly larger than other perturbations in the radiation era~\cite{RC}, so details of the radiation era are critical for a full analysis.
Given such uncertainties in the theoretical parameter estimates, we considered a full likelihood analysis involving different cosmological data to be premature.

We have seen how the CMB constrains the baryon fraction and baryon-photon ratio severely at last scattering (as in FLRW), as well as the area distance to the CMB. Assuming that this can be computed using the LTB solution, this restricts the void profile to be an open model at large distances ($\gtrsim13\,$Gpc say, in comoving distance), but is otherwise broadly in line with SNIa and local Hubble rate measurements. Models which reach asymptotic flatness at the CMB distance can fit, but require a shell around us of low density (for $h^\in\gtrsim0.6$); whether this is at all plausible (i.e., not involving strong fine-tuning in the formation mechanism) remains to be seen. Alternatively,  a high number of relativistic species around today appears to allow asymptotic flatness too (in line with~\cite{huntsarkar}). Of course, much more freedom is possible if the bang time is inhomogeneous~\cite{CFZ}, and it is an important question exactly how this further opens up the possibilities for void models which fit the CMB.
We do not discuss this possibility in detail, taking a conservative approach by assuming variations in the bang-time much larger than time of inhomogeneity creation to be unlikely.

We have argued that the final constraint from the CMB, Eq.~(\ref{glue}), is more nebulous, since it is a constraint on the primordial radiation density profile (or, alternatively, the bang time function). We argue in the Appendix that a radiation density profile with a similar FWHM to the matter and an inhomogeneity $\mathcal{O}(1)$ around decoupling, which gives the correct temperature in and out that we calculate using our FLRW approximations, will easily satisfy this constraint.  However, given that the full two/three-fluid solution is lacking, this requires further investigation to quantify properly. 

Is it physically reasonable that small changes to the details of the radiation era along the central worldline and asymptotically can lead to $\mathcal{O}(1)$ changes to the model today? Yes. Even though the radiation free-streams after decoupling, the imprint it leaves in the dynamics at early times is very important. At equality, if the Hubble rate at the centre and asymptotically are even a little bit different, then this difference in Hubble rate will typically grow when measured on surfaces of constant time (unless the curvature is constant). (We discuss this in the Appendix (see Fig.~\ref{Ht_evol}).) Conversely, then, evolving an inhomogeneous model backwards from today (if we knew how to do it) results in mild inhomogeneity at early times.

In other words, we found that voids including a non-adiabatic mode, i.e., an $\mathcal{O}(1)$ isocurvature mode at early times, can fit CMB data much more easily with respect to the pure adiabatic cases considered in the literature so far.

A number of other issues can now be further understood, in light of our analysis, where properties of the CMB are used for constraining cosmological models. Let us consider briefly the BAO and measurements of the CMB anisotropies as seen by distant observers.\\

\noindent{\it\textbf{Baryon Acoustic Oscillations}}: In principle, the BAO are a smoking gun test of inhomogeneous models because they provide a measurement of $H(z)$ independently of distance measurements~\cite{CBL,gbh1,ZMS}. An LTB model which fits the SNIa data will have a different $H(z)$ from a dark energy FLRW model with the same distance modulus (unless the bang time function is fine-tuned~\cite{CBKH}). However, the BAO rely on comparing the size of the sound horizon at the baryon drag epoch with the same feature imprinted in the galaxy power spectrum. Irrespective of the details of structure formation, without which no fully reliable predictions can be made, this method suffers a serious drawback in void models in directly determining $H(z)$. This is because the sound horizon at the baryon drag epoch need not be spatially constant. The BAO technique measures (see e.g.,~\cite{Percival:2009xn}), from the spherically averaged BAO peak positions at redshift $z$,  $r_s(a_d)/D_V(z)$, where $D_V(z)$ is the `volume distance' $\propto [d_A(z)^2/H(z)]^{1/3}$, and  $r_s(a_d)$ is the comoving sound horizon at the baryon drag epoch~$a_d$ \emph{along the timelike worldline of an observer seen at $z$}, which is inside our past lightcone. In FLRW this is just the same as measured in the CMB angular power spectrum (strictly speaking, it's inferred from it since $z_d<z_\dec$). In LTB, this is an extra degree of freedom which must be measured separately, unless it is assumed. Even taking ratios of BAO measurements does not help because the sound horizon parts do not cancel as they do in FLRW. In principle, then, the BAO tell us only about the radial profiles of $f_b(r)$ and $\eta(r)$, and not, unfortunately, $H(z)$. (Of course, to study the BAO in detail structure formation must be properly calculated.)\\

\noindent{\it\textbf{Off-lightcone CMB observations}}: There have been a number of suggestions of how to constrain the dipole and higher anisotropies of the CMB as seen by distant observers~\cite{goodman,Caldwell:2007yu,GarciaBellido:2008gd,Garfinkle:2009uf}. These look for spectral distortions of the CMB from anisotropic scattering~\cite{goodman,Caldwell:2007yu}, or they use the kinematic Sunyaev-Zeldovich effect to measure the peculiar velocity of clusters with respect to the CMB frame~\cite{GarciaBellido:2008gd,Garfinkle:2009uf}. 
These elegant methods rely principally on the large CMB dipole as seen along our past lightcone by observers comoving with the matter, which is proportional to the peculiar velocity between the CMB rest frame and the matter rest frame (see the Appendix for the definition and discussion of this). 
However, because the temperature of decoupling $T_\dec$ has freedom as a function of $r$ from the freedom in $f_b(r)$ and $\eta(r)$, it can be adjusted to make the dipole seen by off centre observers small. We can see this is feasible because the dipole calculated approximately in LTB comparing the redshift looking in, to the redshift looking out, is of order the percent level; the change in $T_*$ possible by fiddling mildly with $f_b$ and $\eta$ is also of order the percent level (see~\cite{Yoo:2010ad} for an example).  As we discuss in the Appendix, when considering the two-fluid solution, to calculate the peculiar velocity we need to know the radiation profile (and a full integration of the field equations). In essence then, \emph{spectral distortions of the CMB and the kSZ effect in void models are a direct measurement of the radiation profile}. While measurements of zero dipole for all off-centre observers would imply unrealistic fine-tuning problems and so effectively rule void models out, it remains to be seen whether sensible profiles and models will be compatible with future observations of these effects. (Note that a systematic dipole measured would be a smoking gun for voids.) This is an important topic for future research, because if we can observationally show that the dipole, quadrupole and octopole of the CMB around distant observers is small, then this is a powerful test of the FLRW models~\cite{CM,Ellis:1983}. 
This is also important to test the adiabaticity of the history of the universe. Indeed the extraction of temperature $T(z)$ is typically degenerate with the degree of isotropy of the sky at redshift $z$ (see, e.g.,~\cite{Luzzi:2009ae} where $T(z)$ is extracted from SZ effect and shows strong correlation with the velocity between cluster and CMB frame which drives the kSZ effect).

\begin{figure}[t]
\begin{center}
\includegraphics[width=\columnwidth]{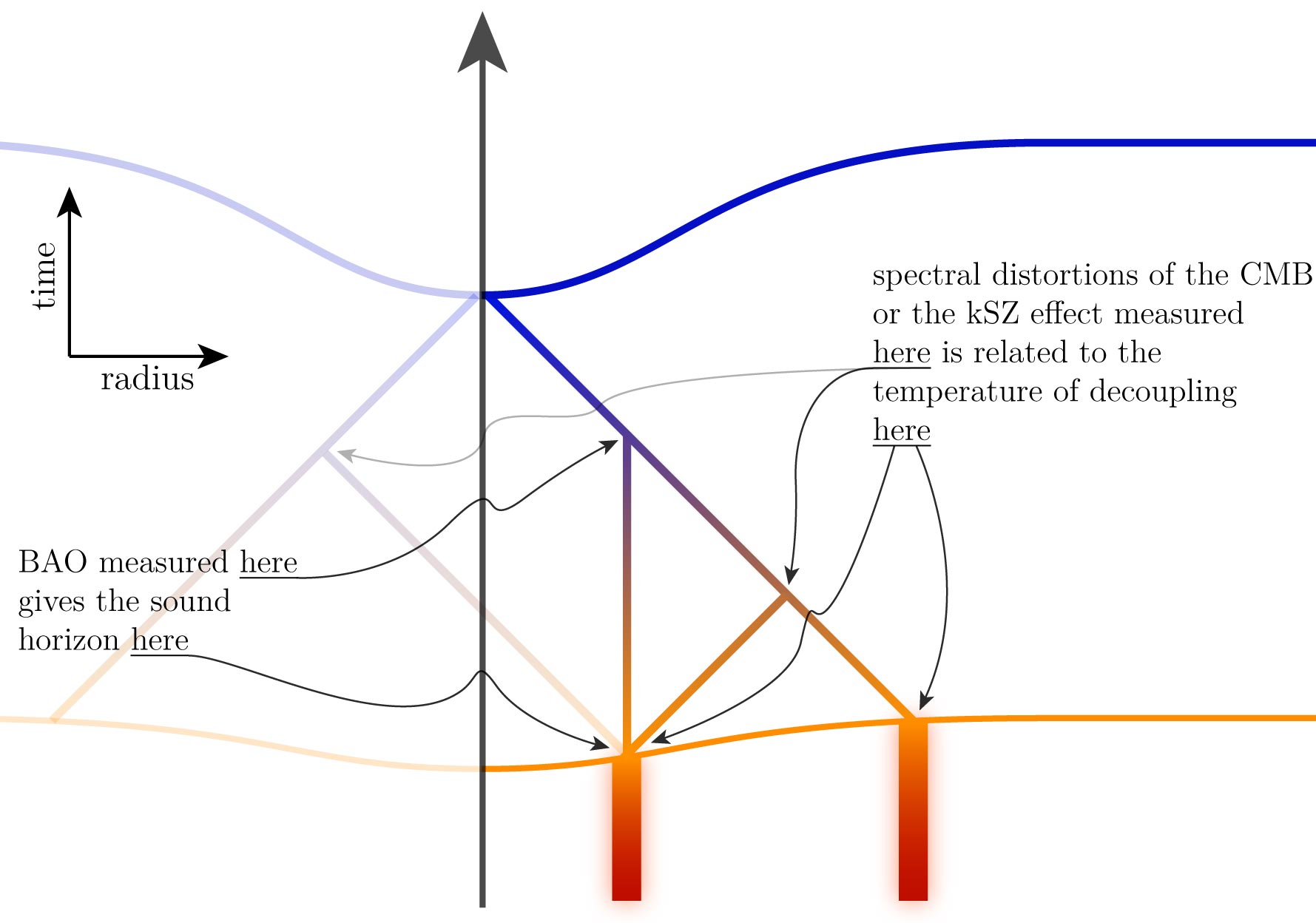}
\caption{BAO and measurements of the CMB anisotropies around distant observers tell us conditions around the time of decoupling inside our past lightcone, and so help us to reconstruct the radiation profile. To use these probes to rule out void models they need to be cross-correlated with other observations in different directions such that they both measure conditions at the same radius.}
\label{fig:bao}
\end{center}
\end{figure}

The future, then, of these `Copernican tests' lies in cross-correlation of measurements, as illustrated in Fig.~\ref{fig:bao}. Given enough observations we can reconstruct the conditions across the entire last scattering surface inside our past lightcone. 

As a final comment, we also note that constraints on our distance from the centre of the void from the CMB~\cite{Alnes:2006pf,mortsell} might change when a more thorough analysis of the peculiar velocity is undertaken using the field equations presented in the Appendix. For example, since the peculiar velocity depends on the radiation profile one can presumably construct models such that distant observers see no dipole at all. In fact, this mechanism could be used to construct void models in which the Copernican problem is dramatically relieved (at the expense of a temporal coincidence problem). Higher multipoles can then be used to place constraints on the distance from the centre, but would lead to much weaker bounds.

\section{Conclusion}
\label{sec:conc}

The ethos of modeling the universe with an inhomogeneous model is different from the standard FLRW models.  Even where FLRW models contain quintessence fluids or modified gravity, flat $\Lambda$CDM is the simplest model common to all possibilities and serves as a fixed point from which to smoothly interpolate from. Extensions to the concordance model typically raise more questions than they answer. This means that, while $\Lambda$CDM is a good fit to the data, Occam's razor and Bayesian Information Criteria can be used to good effect. If it ain't broke don't fix it, so to speak. 

But if we consider the universe inhomogeneous on Hubble scales (and larger), then in some sense we must aim to reconstruct the density and curvature distributions etc., directly from observations, rather as if we were making a map. A priori assumptions such as asymptotic flatness or Gaussian profiles, however reasonable, are basically arbitrary. This is unfortunate in a sense because it gives us far too much freedom for model building for our fairly limited observations (even in the idealized case~\cite{CM}). An exception to this would come if we had early universe models for the creation of a Hubble scale void~-- such a model would presumably come with restrictions on the possibilities available, and so some hope of ruling them out.\footnote{This is not as far fetched as it might seem: there are a number of speculative ideas which can create large spherical features that exist today, e.g.,~\cite{Fialkov:2009xm,Kovetz:2010kv,Afshordi:2010wn}.} 

In this context, then, the CMB is not as restrictive as it might first appear. Indeed, we find it much less restrictive than previous works which neglect the dynamical contribution from the radiation. Clearly it tells us that the universe must be nearly spherically symmetric, and, in the simplest interpretation, that we must be fairly close to the centre. Other than that, we have seen that it tells us the baryon fraction and baryon-photon ratio in one sphere around us, as well as the area distance to the CMB. This is just as in FLRW. Our main finding is that, given a monotonic void profile and a void which fits the local Hubble rate and SNIa data, a void which is open with $\Omega_m^\out\sim0.7$ will fit the CMB without any problems, as already anticipated in~\cite{RC}.  In particular, we find that the requirement in other analyses~\cite{ZMS,MZS} that the CMB enforces a low $H_0$ locally is an artifact of attempting to model the full spacetime of inhomogeneous matter and radiation a a separate LTB dust model embedded in radiation filled FLRW. Rather, we have argued that the full solution of spherically symmetric radiation plus matter is required before we can say what the precise constraints on void models of dark energy are.

\acknowledgments

Special thanks to Timothy Clifton and Roy Maartens for extensive discussions, and to James Zibin for a vibrant exchange of views. We thank George Ellis and Pedro Ferreira for discussions. This work is supported by the NRF (South Africa), and MR acknowledges funding by the Centre for High Performance Computing, Cape Town.

\appendix

\section{The importance of the radiation}
\label{sec:rad}
A key part of our approach lies in including the radiation energy density. We are able to do this only because the spacetime is asymptotically FLRW; a comprehensive analysis of the CMB in a spherically symmetric model including radiation is a significant challenge. But why is it really important to include the radiation?

In essence, it follows from the fact that $$a_\dec^\out=T_0^\out/T_\dec^\out$$ can be significantly different from $1/(1+z_\dec)=T_0^\in/T_\dec^\out$, if $T_0^\out$ is significantly different from $T_0^\in$. This important effect is missed when the dynamics of radiation is ignored.  
Indeed, in a pure dust LTB model, one may calculate $a_\dec$ by choosing a void model, integrating out to $z_*$ using an LTB solution, and then calculating $a_*$ at that redshift. A key feature of the LTB solution for the types of voids considered here is that the function $a(z)$ one finds is very close the the FLRW relation $a(z)=1/(1+z)$, with a difference of a few percent. Now, since the spacetime is asymptotically FLRW, the temperature asymptotically today must be given by $T_0^\out=T_\dec^\out/a_\dec^\out$~-- the standard FLRW relation. In this case $T_0^\out$ and $T_0^\in$ could only differ by a few percent. But this seems to contradict Eq.~(\ref{varpi_b}), which also must hold asymptotically, and fixes $T_0^\out$ given $f_b$ and $\eta$ from the CMB, and $\Omega_mh^2$ from the void shape. As we show in Sec.~\ref{sec:result}, this gives a difference between $T_0^\out$ and $T_0^\in$ of tens of percent, not just a few.

This makes sense: the CMB is emitted at nearly constant temperature. At the centre of the void the Hubble rate is much higher than it is asymptotically, so it would seem natural that the temperature on a surface of constant time should be lower at the centre because it is moving away from the observed patch of the LSS faster in all directions.
Moreover, provided that the central and asymptotic worldlines are (very close to) FLRW and that $T\propto(1+z)$ along null cones, a difference between $T_0^\out$ and $T_0^\in$ can be simply translated into a constraint for the radiation profile at a time around $t_\dec$. Tens of percent give an $\mathcal{O}(1)$ inhomogeneity in the density, which is comparable to the matter void.

\begin{figure}[ht]
\begin{center}
\includegraphics[width=\columnwidth]{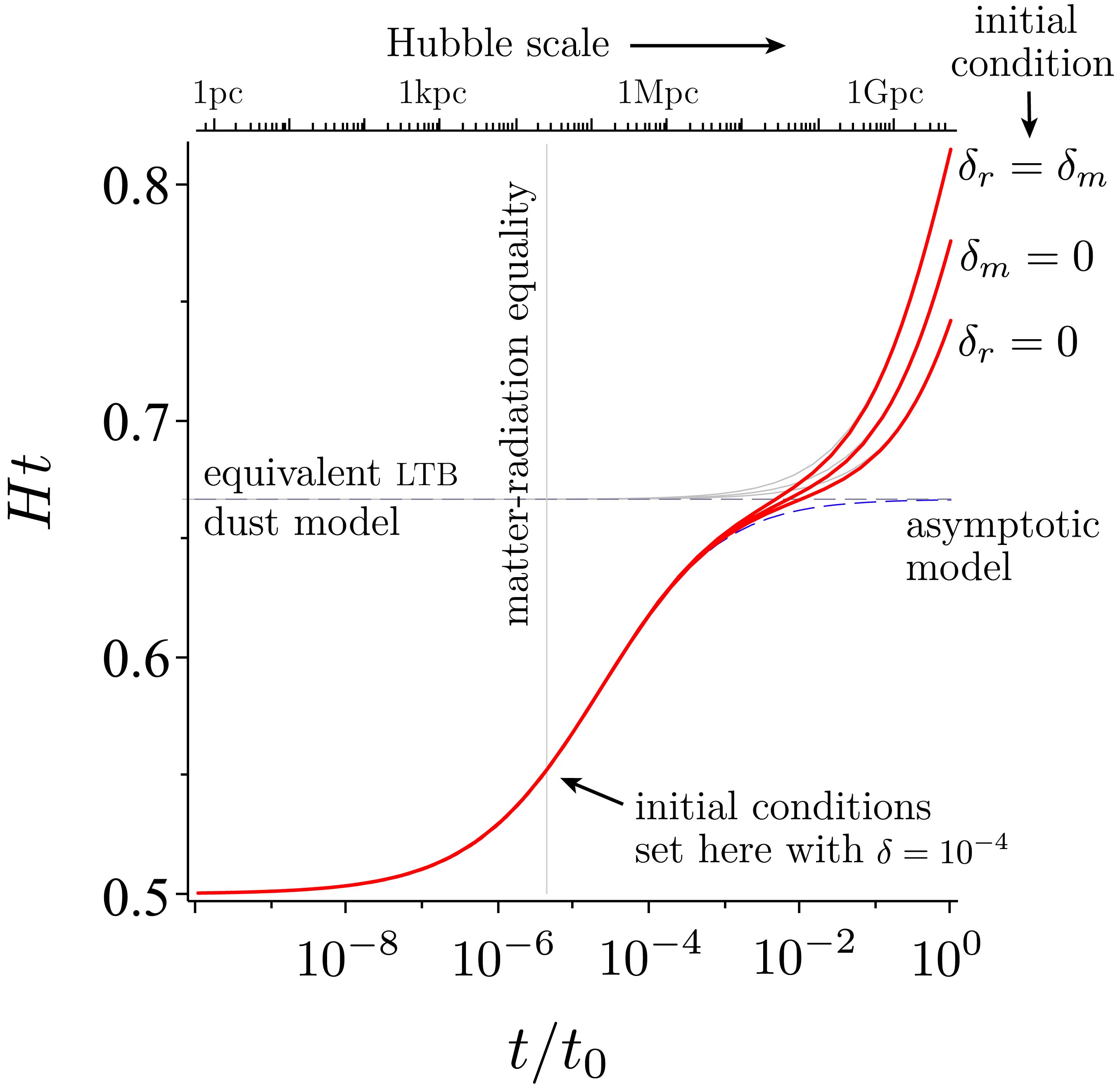}
\caption{Evolution of the Hubble rate for models set with different initial conditions at $t_\eq^\out$. We have chosen $\delta=10^{-4}$ and $h=10^5$, and considered different cases: inhomogeneity in the matter only, inhomogeneity in the radiation only, and equal inhomogeneity in the radiation and matter. Clearly these lead to very different behavior at late times. We have set $t_0=12\,$Gyr. Note that we can expect this FLRW evolution to be accurate while the void is larger than the Hubble scale~-- that is, until the Hubble scale reaches 1\,Gpc or thereabouts. After than, we might expect the radiation temperature to evolve differently, but not the overall dynamics shown here.}
\label{Ht_evol}
\end{center}
\end{figure}

\begin{figure}[ht]
\begin{center}
\includegraphics[width=\columnwidth]{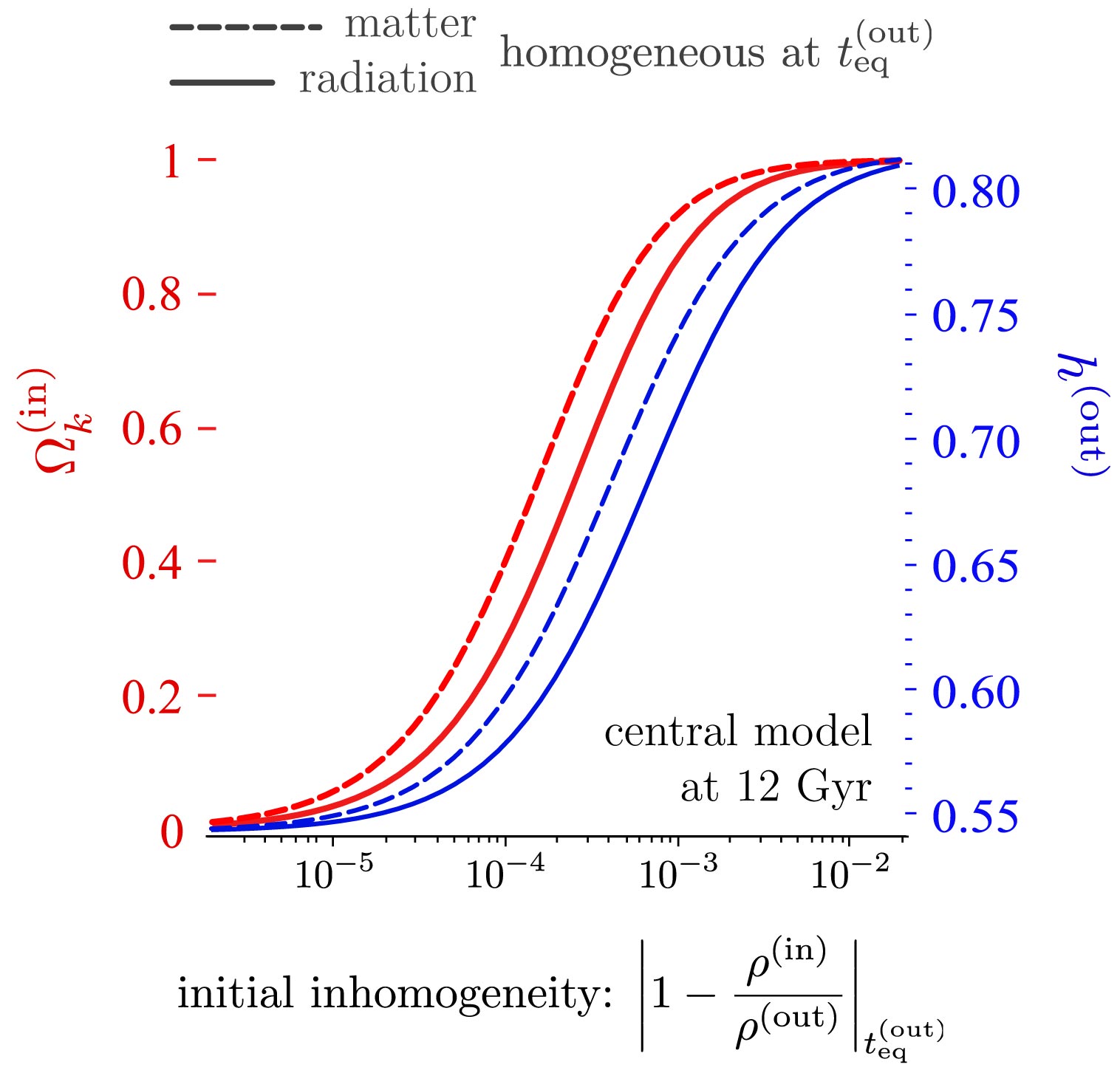}
\caption{Central curvature and Hubble rate at 12\,Gyr for an asymptotically flat model, as a function of initial inhomogeneity which is set at matter-radiation equality. The dashed curves have homogeneous matter content initially, and so the horizontal axis represents the inhomogeneity in the radiation; the solid curves represent the reverse situation. We have a significantly different curvature at late times depending on whether the matter or radiation are chosen homogeneous. Obviously even larger differences appear for the generic case when both matter and radiation are inhomogeneous (not shown).}
\label{inhomog_evol}
\end{center}
\end{figure}

Another way to see this is to consider what is going on at early times. Could a small difference in the energy density of the radiation at the centre compared to asymptotically really make much of a difference to later evolution? Perhaps surprisingly, yes. Let us illustrate this as follows.\footnote{While the scale of the void is much larger than the Hubble scale, we can calculate the evolution of the parameters of the centre of the void as in FLRW~-- since the void is Gpc, and the Hubble scale today is $\sim 4\,$Gpc, we can estimate that FLRW evolution should be good up until, say, $0.1-1\,$Gyr. By this time, radiation is subdominant and makes sub-percent changes to the evolution history. Provided we are interested only in the overall dynamics, and not the radiation energy density (which may have corrections to FLRW evolution~-- see below), we can calculate the future evolution as in FLRW.}  Consider starting the evolution of a void model from matter-radiation equality to late times, and let us assume an asymptotically flat model for definiteness. Asymptotically at matter-radiation equality, $t_\eq^\out$, choose $h^\out(t_\eq^\out)=10^5$; evaluated at 12\,Gyr this model has $h\approx0.54$ and $\Omega_m\approx0.9996$. Now set the initial conditions at the centre \emph{at the same cosmic time} $t_\eq^\out$; we fix the density parameters and tune $h$ to make this precise. Unless the model is flat at the centre too this time will not be quite the central value of matter-radiation equality, but it will be close to it. We find significantly different future evolution depending on how set the initial conditions on the matter density and radiation density. That is, if we compare a model with homogeneous radiation density at $t_\eq^\out$ we find a different model from choosing the matter density homogeneous by a significant amount at late times. This difference does not just decay away, as we see in Figs.~\ref{Ht_evol} and~\ref{inhomog_evol}, but grows to an $\mathcal{O}(1)$ difference in the curvature and Hubble rate. This is because at a fixed time homogeneous matter vs radiation require different Hubble rates, which translates to different curvatures; differences in curvature is important. (Note that if we run the model backwards the homogeneous component does not stay homogeneous, of course, so these fine-tuned examples are just for illustration.)

Hence, radiation must be important when computing relations at high redshift as we now discuss. This features in the constraint given by Eq.~(\ref{glue}).

\subsection{Using matching to understand the unexpected sensitivity of including radiation}
\label{apprad}

\begin{figure}[htbp]
\begin{center}
\includegraphics[width=\columnwidth]{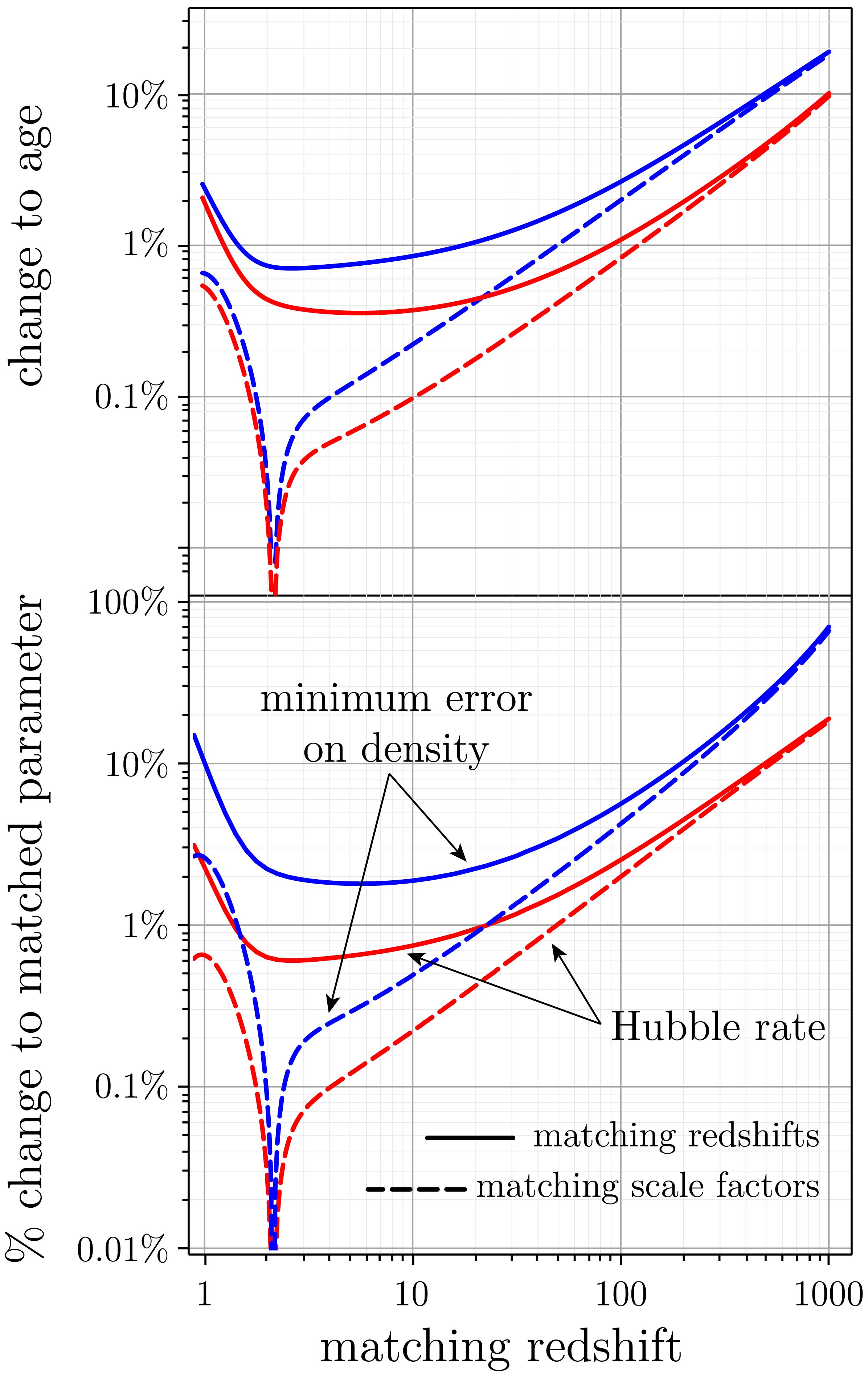}
\caption{Matching a dust LTB model which is asymptotically flat, with $h^\in=0.65, \omin=0.2$ and FWHM$\,=3\,$Gpc, to a radiation filled FLRW at some intermediate redshift $z_m$.  Even though the radiation is sub-dominant after decoupling, this matching incurs $\mathcal{O}(\%)$ level errors to the parameters from the asymptotic LTB model, which translates to large errors on the age (top). This is far too high for the accuracy required to determine $z_\dec^\in/z_\dec^\out=T_0^\out/T_0^\in$ correctly. Matching is performed such that the Hubble rates agree at $z_m$ in the LTB model: $H_\|^{\text{LTB}}(z_m)=H^{\text{FL}}$, but we have to make the choice of exactly how to match up to the FLRW spacetime. We try two choices for where to match $H^{\text{FL}}$, at $z=z_m$ and $a=a_\|^{\text{LTB}}(z_m)$; even this minor change gives large changes in model parameters.}
\label{fig:error}
\end{center}
\end{figure}

The main difference in our approach from~\cite{ZMS} and others is that we relieve a constraint on $T_0$ spatially which translates into a very low $h^\in$. We can see schematically why this constraint cannot be used by considering matching an LTB model to an FLRW with radiation at some intermediate redshift $z_m$. It is assumed in~\cite{ZMS} that we can choose $z_m$ high enough to be outside the void, but low enough for radiation to not be important.  However, matching a dust model to a radiation filled FLRW model typically brings in percent changes to the asymptotic parameter values, as illustrated in Fig.~\ref{fig:error}; these are significant. 

Consider the constraint Eq.~(\ref{glue}), which takes us from the top of the central worldline today to the point of last scattering, as seen from the centre, along the null cone. This is the constraint used in other works to show that models with radiation must have $T_0^\in\simeq T_0^\out$. Here we will demonstrate that, without knowing the full spacetime including radiation everywhere, it is far too sensitive to the radiation era to use in any approximate manner.

We can write Eq.~(\ref{glue}) approximately as 
\be
t_0-t_\dec^\out=-\int_0^{z^\in_{\dec}}\frac{d z}{(1+z)H_\|(z)}\,,\label{glue2}
\ee
where we have assumed a simplified form for this relation for illustration (namely, to follow other work, while actually this integral has additional corrections to $H_\|(z)$ from evolving curvature terms which we discuss below). Here, $H_\|(z)$ represents the real Hubble rate in the spacetime with matter plus radiation along the past null cone of the observer at the centre (which we don't know how to calculate), and $z^\in_{\dec}=T_\dec/T_0^\in-1$ is the redshift of decoupling as seen from the centre. The times on the left are the proper age of the whole spacetime (calculated including radiation) $t_0$, and the proper time of decoupling where it is observed, in the asymptotic part of the spacetime, $t_\dec^\out$. Now, an asymptotic observer will measure the same time difference to last scattering but at a different redshift $z_\dec^\out=T_\dec/T_0^\out-1$, with a Hubble rate along their past lightcone $H^\out(z)$, implying
\be
t_0-t_\dec^\out=-\int_0^{z^\out_{\dec}}\frac{d z}{(1+z)H^\out(z)}\,.\label{glue_out}
\ee

\begin{figure}[t]
\begin{center}
\includegraphics[width=\columnwidth]{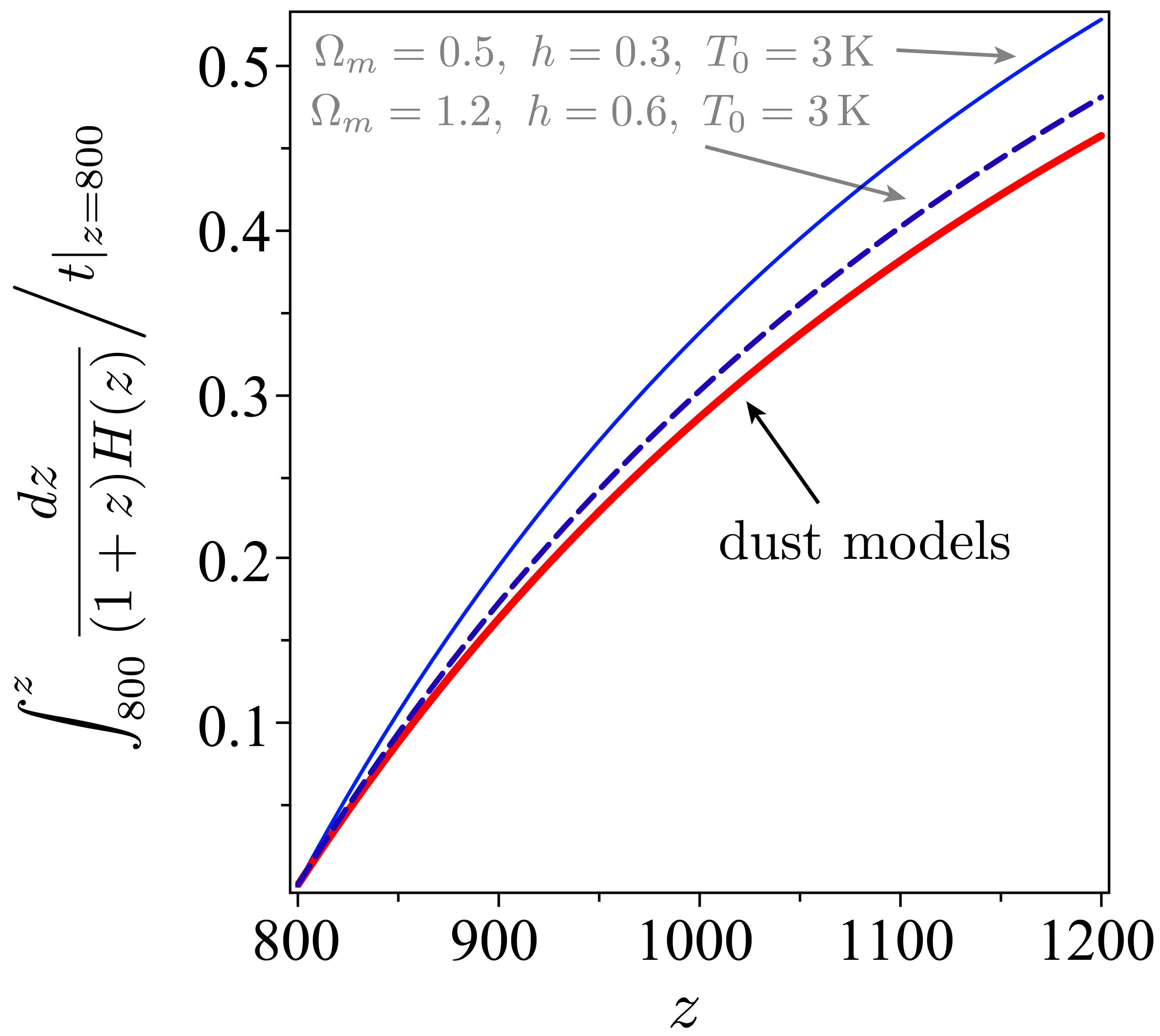}
\caption{The function $t(z)$ is insensitive at high redshift. Uncertainty in this integral by factors of $t_\dec$ result in differences of hundreds in redshift. Here we have shown dust models (all of them), and some extreme radiation + matter models for comparison.}
\label{z_star}
\end{center}
\end{figure}

Now let us assume that we can model the matter and radiation model of the void as a dust model out to some matching redshift $z_m$, and as an FLRW model \emph{with} radiation from $z_m$ to $z_\dec^\in$, ignoring the subtleties involved in the matching. This implies that
\bea
\int_0^{z^\in_{\dec}}\frac{d z}{(1+z)H_\|(z)}&\simeq& \int_0^{z_m}\frac{d z}{(1+z)H_\|^{\text{LTB}}(z)}\nonumber\\&+&\int_{z_m}^{z^\in_{\dec}}\!\!\!\frac{d z}{(1+z)H^\out(z)}\,.
\eea
The first of these integrals is the difference in time at the centre of an equivalent LTB dust model and the time at the matching point in the same model, viz.:
\be
-\int_0^{z_m}\frac{d z}{(1+z)H_\|^{\text{LTB}}(z)}=t_0^{\text{LTB}}-t_m^{\text{LTB}}\,.
\ee
The second integral may be formally split at $z_\dec^\out$:
\bea
\int_{z_m}^{z^\in_{\dec}}&&\!\!\!\!\!\!\!\!\!\frac{d z}{(1+z)H^\out(z)}=\\ &&\left\{\int_{z_m}^{z^\out_{\dec}}+\int_{z_\dec^\out}^{z^\in_{\dec}}\right\}\frac{d z}{(1+z)H^\out(z)}\nonumber\,.
\eea
Now, in Eq.~(\ref{glue_out}) we can formally split the integral at the same $z_m$, even though it has no significance asymptotically,
\bea
\int_{0}^{z^\out_{\dec}}&&\!\!\!\!\!\!\!\!\!\!\!\!\!\!\!\!\!\frac{d z}{(1+z)H^\out(z)}=\nonumber\\ &&\left\{\int_{0}^{z_m}+\int_{z_m}^{z^\out_{\dec}}\right\}\frac{d z}{(1+z)H^\out(z)}\nonumber\,\\
&=&-t_0+t_m^\out+\int_{z_m}^{z^\out_{\dec}}\!\!\!\!\frac{d z}{(1+z)H^\out(z)}\,.
\eea
Gathering these results together, Eq.~(\ref{glue2}) may now be written as
\bea\label{key}
\left(t_0^{\text{LTB}}-t_0\right)-\bigl(t_m^{\text{LTB}}-t_m^\out\bigr)=~~~~\nonumber\\
\int_{z_\dec^\out}^{z_\dec^\in}\frac{d z}{(1+z)H^\out(z)}\,.
\eea
The integral on the rhs will be $\mathcal{O}(t_\dec)$ even when there is a difference of hundreds in the integration limits, because it is at such high redshift~-- see Fig.~\ref{z_star}.
This is a key result: we can expect a large difference in ${z_\dec^\out}$ compared to ${z_\dec^\in}$, and hence a large difference between $T_0^\in$ and $T_0^\out$, even if the difference on the left hand side of this equation is $\mathcal{O}(t_\dec)$ (which is a tiny fraction of $t_0$).  In general we expect 
\be
\left(t_0^{\text{LTB}}-t_0\right)\sim t_\dec
\ee
since the age of the LTB model, which does not include radiation, is different from the correct age by a significant multiple of the decoupling time along the central worldline (typically tens of times the decoupling time). Similarly, the difference in time at the matching redshift between the LTB dust and the matched radiation model, $\bigl(t_m^{\text{LTB}}-t_m^\out\bigr)$, must also be $\mathcal{O}(t_\dec)$. This implies that we expect, generically, 
\be
\int_{z_\dec^\out}^{z_\dec^\in}\frac{d z}{(1+z)H^\out(z)}\propto t_\dec
\ee
with factors of 10 involved in the proportionality. This implies a difference in redshift of hundreds in the decoupling redshift between central and asymptotic observers, which is in the tens of percent~-- exactly what we find in $T_0^\in$ vs. $T_0^\out$. What this also implies is that it is inevitable to find differences $\mathcal{O}(10\%)$ in $z_\dec$ depending on the precise nature of the matching conditions and the LTB model used~-- even though the overall errors appear to be tiny. Indeed, it is not too surprising to note that when defining the LTB model at the centre, which approximates the actual spacetime with radiation we are trying to model, there is further ambiguity: when we remove the radiation, do we keep the matter density fixed or the curvature fixed (or the age, or $h$)? It sounds irrelevant but this also leads to differences in $\left(t_0^{\text{LTB}}-t_0\right)$ of $\mathcal{O}(t_\dec)$, and consequently tens of percent in $z_\dec$.

Is there any way around Eq.~(\ref{key})? For nearly homogeneous radiation today, as found in~\cite{ZMS} (the low $h^\in$ condition), we need the left hand side to vanish, or at least be much smaller than $t_\dec$. Let's say we adjust our parameters in the LTB model such that $t_0^{\text{LTB}}-t_0=0$; then, can we adjust the matching time or any aspect of the void to set $t_m^{\text{LTB}}-t_m=0$? It does not appear so because a dust and radiation FLRW do not stay synchronised. Similarly, one can fine tune a matching redshift such that $\left(t_0^{\text{LTB}}-t_0\right)=\bigl(t_m^{\text{LTB}}-t_m^\out\bigr)$, (effectively, by adding an error associated with the curvature which cancels the error on radiation; e.g., a model with $\Omega_m^\in=0.5,\omout=0.7,h^\in=0.6$ does this if we choose $z_m\approx2.7$, and are careful to match the LTB scale factor to the FLRW one at $z_m$; alternatively, a model with $\Omega_m^\in=0.4,\omout=0.7,h^\in=0.6$ matched at $z=z_m\approx6.95$ can achieve this), but this actually then implies the wrong temperatures in and out today; so this is no measure of accuracy.
To avoid complications associated with curvature one has to choose a matching redshift at $z_m\gg2-3$, as also shown in Fig.~\ref{fig:error}. In this regime, the error introduced by the approximated treatment of radiation in the matching procedure grows with redshift, and gives changes much larger than $t_\dec$ in general. 

So even though the radiation is a very small contribution to the overall energy density at $z_m$, it makes a vital difference to a model with matter and radiation when we integrate to high redshift.  This is because we have to be able to synchronise the time coordinates asymptotically at very early times with those used at the centre at very late times to each other in a robust way. It appears that matching a dust LTB model to a radiation filled FLRW model is subtle and cannot be sensitive enough for the results we are interested in.

\subsection{Lemaitre solution for one fluid}
\label{sec:onefluid}
An inhomogeneous, spherically symmetric Universe can be described by the Lemaitre metric~\cite{Lemaitre:1933gd}:
\be
ds^2=-e^{2\,\phi (t,r)}dt^2+\frac{a_\parallel^2(t,r)}{1-k(t,r)r^2}dr^2+a_\perp^2(t,r)\,r^2 d\Omega^2
\ee
where $a_\parallel \equiv(a_\perp r)'$, $'={\partial\over\partial r}$ and we will use $\dot{}$ $={\partial\over\partial t}$ .
The time coordinate $t$ is the proper time for an observer at $r_0$ having defined $\phi (t,r_0)=1$.

For a universe filled by one perfect fluid with equation of state $p=w \rho$, the Einstein equations can be recast into (see also~\cite{Alfedeel:2009ef} for analogous equations with different notation):
\bea
\tilde{H}_\perp^2 &=& \frac{8\pi G}{3} \tilde{\rho}(t,r)-\frac{k(t,r)}{a_\perp^2}+\frac{\Lambda}{3}\label{eq:Friedmann}\\
H_\perp&\equiv& \frac{\dot a_\perp}{a_\perp}=\tilde{H}_\perp \,e^{\phi}\label{eq:def}\\
\rho &=& \tilde{\rho}+\tilde{\rho}' \frac{a_\perp r}{3 a_\parallel}\;,\\
\dot {\tilde{\rho}} &=& -3\frac{\dot a_\perp}{a_\perp}(1+w)\tilde{\rho}-w \frac{\dot a_\perp r}{3 a_\parallel}\tilde \rho'\label{eq:rho_evol}\\
\dot k &=& -2\phi'\frac{(1-k\,r^2)}{r} \frac{\dot a_\perp}{a_\parallel}\label{eq:k_evol}\\
\phi' &=& -\frac{p'}{\rho+p}\label{eq:sigma}
\label{cLres}
\eea

In case of pure dust, $w_m=0$ and the LTB solution is recovered.
Indeed, Eq.~(\ref{eq:rho_evol}) implies $\tilde{\rho}\propto a_\perp^{-3}$, so that the time-dependence of $\tilde{\rho}$ is entirely encoded in $a_\perp$, and the mass $M=8\pi G/3\, \tilde{\rho}\, a_\perp^3$ inside a comoving shell of radius $r$ is conserved.
Moreover, zero (or homogeneous) pressure leads to $\phi'=0$ in Eq.~(\ref{eq:sigma}) and, by a simple change of time coordinate, to $\phi=0$. This in turn means $k=k(r)$, see Eq.~(\ref{eq:k_evol}), and the Friedmann equation can be rewritten in the standard way along each wordline but in terms of $\tilde{\rho}$ instead of $\rho$.

The picture is different for a radiation fluid, where $w_r=1/3$ and the evolution of $\tilde{\rho}$ deviates from the $a_\perp^{-4}$ scaling. Indeed, the inhomogeneous pressure leads to exchange of entropy between neighbourhood regions and so to non-conservation of entropy inside comoving shells. Because of this, the curvature along a given worldline no longer stays constant as it does in an LTB universe, but rather its time-evolution is related to the radiation inhomogeneity via Eqs.~(\ref{eq:k_evol}) and~(\ref{eq:sigma}). This feature is clearly present also in the general case of an inhomogeneous universe including both matter and radiation components and could lead, in the matter era, to a non-negligible deviation from LTB dynamics depending on radiation profile, as we will illustrate.

\subsection{Einstein equations including matter and radiation fluids}
\label{sec:efe}
Now we consider a universe filled by matter and radiation which have energy momentum tensors:
\bea
T^{ab}_{m}&=& \rho n^a n^b\\
T^{ab}_{r}&=& \frac{4}{3}\mu u^a u^b+\frac{1}{3}\mu g_{ab}\;,
\eea
where $\rho$ and $n^a$ are energy density and four-velocity of matter, $\mu$ and $u^a$ are energy density and four-velocity of radiation, and $g_{ab}$ is the metric tensor. We have neglected the anisotropic pressure term for the radiation which couples to the hierarchy of multipole equations from the Liouville equation.
Note that, in an inhomogeneous universe, matter and radiation are not comoving, and so $n^a \neq u^a$.
For the large and shallow void models considered here, the peculiar velocity $v_p$ between the two frames is however expected to be small until late times, namely, until the inhomogeneity crosses the horizon.
We describe baryons and dark matter as a single matter fluid, which is an approximation since baryons are coupled to photons until last scattering. 
However, the peculiar velocity at LSS is typically extremely small and so these two matter fluids can be confidently treated as comoving. 
In the following we introduce the 3-velocity $v^a$ (where the peculiar velocity is $v_p=\sqrt{g_{ab}v^a v^b}$) and present the Einstein equations to first order in the peculiar velocity, i.e., neglecting terms $\mathcal{O}(v_p^2)$ in the solution. (The full equations may be written down but they are seemingly intractable.)

We have a choice to make for the coordinates. We consider two cases: a frame comoving with matter and a frame comoving with radiation. 

\noindent{\bf Matter frame}

The coordinates are fixed such that 
\bea
n^a&=&(e^{-\phi},0,0,0),\nonumber\\
v^a&=&(0,v,0,0)\nonumber\\
u^a&=&\gamma(n^a+v^a)\,:~\gamma=1/\sqrt{1-v_av^a},~n^av_a=0. \nonumber
\eea
where $v=v(t,r)$. 
In this case, $\nabla_aT^{ab}_m=0$ implies that $\phi'=0$, and so we can set $\phi=0$ as a gauge degree of freedom. The equations of motion are then:
\bea
H_\perp^2&=&\frac{2M}{a_\perp^3r^3}-\frac{k}{a_\perp^2}\nonumber\\
\dot M &=& -\frac{2}{3}a_\|a_\perp^2r^2\mu v-\frac{1}{6}\mu r^3 a_\perp^3 H_\perp\nonumber\\
\dot k &=& -\frac{4 a_\| a_\perp }{3r} \mu v\nonumber\\
\dot \rho &=& -(H_\|+2H_\perp)\rho +
\frac{2 a_\|a_\perp r}{3(1-kr^2)} \rho\mu v\nonumber\\
\dot v &=&\frac{1}{3}(2H_\perp-5H_\|) v -
\frac{(1-kr^2)\mu'}{4 a_\|^2 \mu}\nonumber\\
\mu&=&\frac{(a_\|a_\perp^2 r^2\rho  -2 M')(4r a_\| a_\perp  H_\perp  v -3(1-k r^2))}{3a_\|a_\perp^2r^2(1-kr^2)\nonumber}
\eea

\noindent{\bf Radiation frame}

The coordinates are fixed such that 
\bea
u^a&=&(e^{-\phi},0,0,0),\nonumber\\
v_a&=&(0,v,0,0) \nonumber\\
n^a&=&\gamma(u^a+v^a)\,:~\gamma=1/\sqrt{1-v_av^a},~n^av_a=0\,,\nonumber
\eea
(note the different definition for $v$) which leads to:
\bea
\tilde H_\perp^2&=&\frac{2M}{a_\perp^3r^3}-\frac{k}{a_\perp^2}\nonumber\\
\dot M &=& -\frac{a_\perp^2r^2e^\phi(1-kr^2)}{2a_\|}\rho v-\frac{1}{6}\mu r^3 a_\perp^3 H_\perp\nonumber\\
\dot k &=& -\frac{a_\perp e^\phi(1-kr^2)}{a_\|r}\rho v+\frac{a_\perp H_\perp(1-kr^2)\mu'}{2a_\|r\mu}\nonumber\\
\dot\mu &=& -\frac{4}{3}(H_\|+2H_\perp)\mu -\frac{ra_\perp H_\perp }{3a_\|}\mu'+\frac{2a_\perp re^\phi}{3a_\|}\rho\mu v\nonumber\\
\dot v &=& \frac{e^\phi\mu'}{4\mu}\nonumber\\
\phi'&=& -\frac{\mu'}{4\mu}\nonumber\\
\rho &=& \frac{(a_\|a_\perp^2 r^2\mu  -2 M')(ra_\perp  H_\perp e^{-\phi} v -a_\|)}{a_\|^2a_\perp^2r^2\nonumber}
\eea

Note the different meanings of all the variables in the two frames. In the matter frame the main effect of the radiation lies in the evolution of the spatial curvature, while in the radiation frame there are additional effects such as the de-synchronization of clocks between neighboring worldliness directly induced by gradients in the radiation energy density (i.e., $t$ is not the proper time). Note also that the void profile~-- while fixed in LTB~-- can now change in (co-moving) size. Hence, the scale of the radiation profile will tend to grow relative to the matter profile.

In order to solve the field equations, we note that we have a first-order pde system in 5 variables (in the matter frame this is $a_\perp,M,kr^2,\rho,v$; in the radiation frame, we can integrate for $\phi$ trivially and use $a_\perp,M,kr^2,\mu,v$). Given 5 evolution equations we must specify initial conditions for the 5 variables, as well as 5 boundary conditions. The important point for this paper is that the boundary is $r=0$, and the obvious choice of boundary condition is $v=0,\rho'=0,\mu'=0$. This ensures that $\dot M=(kr^2\dot)=0$ along $r=0$, and that this worldline is FLRW. The other boundary is $r=\infty$ and is never affected by $r=0$ (except at $t=\infty$); if the initial conditions are chosen to be flat as $r\to\infty$ then that will give a model which has the same asymptotics as we have used here. Although it might seem intuitive that information from the initial data can propagate into the central worldline and prevent FLRW evolution, the boundary condition there propagates out into the spacetime to ensure this does not happen.

For our purposes, the area distance to LSS in the only quantity for which a full integration of such systems of partial differential equations is needed.
Indeed, as mentioned many times in the paper, we compute the small-scale CMB by splitting the physics of decoupling and line-of-sight effects.
For the first, equations reduce to the FLRW case since at LSS ($r\gg$ void size) gradients are negligible (for Gaussian-like density profiles). The area distance $d_A$ then accounts for line-of-sight effects. 

We motivate below that, for this quantity, LTB formulas can provide a good approximation and, in the rest of the paper, we assumed $d_A(z)$ as computed in the LTB framework.

\subsection{Redshift relations for the central observer}

An incoming photon along a radial null geodesic satisfies $ds^2=0=d\Omega$, which leads to (in terms of the proper time $t$ for an observer located at $r_0$ where $\phi (t,r_0)=1$):
\be
dt=-\frac{a_\parallel(t,r)}{\sqrt{1-k(t,r)r^2}}e^{-\phi (t,r)}dr\;.
\ee
We define $\chi \equiv e^{-\phi}\,a_\parallel/\sqrt{1-k\,r^2}$.
Considering another light ray emitted with an infinitesimal time interval $\tau$ of delay: $d(t+\tau)=-dr\,\chi(t+\tau,r)$, and taking the first order in the Taylor expansion $\chi(t+\tau)=\chi(t)+\dot \chi(t)\,\tau$, one can easily derive $d\tau=-\dot \chi(t)\,\tau \,dr$. Then, from the definition of redshift: $\tau_{obs}/\tau=1+z$, it follows $d\tau/\tau=-dz/(1+z)$ and (see also Ref.~\cite{Lasky:2010vn}):
\bea
\frac{dz}{1+z}&=&\dot \chi(t,r)\,dr\;,\label{dzdr}\\
\frac{dz}{1+z}&=&-\frac{\dot \chi(t,r)}{ \chi(t,r)}dt\;.
\eea
In the LTB case, $\chi_{LTB}=a_\parallel(t,r)/\sqrt{1-k(r)\,r^2}$ and $\dot \chi_{LTB}/\chi_{LTB}=H_\parallel$.
It implies that, along the same geodesic, $1/a_\perp$ scales roughly as $1+z$. Indeed $d \log{(1+z)}/d \log{a_\perp}=dz/dt\cdot dt/da_\perp\cdot a_\perp/(1+z)= -H_\parallel\,\hat H_\perp^{-1}$, which is $\simeq -1$ for smooth shallow voids, where we defined $\hat H_\perp\equiv (\dot a_\perp+a_\perp'\,dr/dt)/a_\perp=H_\perp+(1-a_\|/a_\perp)/(r\cdot \chi)\simeq H_\perp$.

In the more general case considered here this relation becomes:
\bea
\frac{d \log{(1+z)}}{d \log{a_\perp}}&=& -\frac{\dot \chi}{\chi\hat H_\perp}\label{Eq:avsz}\\
&=&-\frac{H_\parallel}{\hat H_\perp}\left[1-\frac{\dot \phi}{H_\parallel}+\frac{\dot k\,r^2}{2\,(1-k\,r^2)\,H_\parallel}\right]\;.\nonumber
\eea
Defining $G\equiv -H_\parallel/\hat H_\perp$ and $F\equiv-\dot \phi/H_\|+\dot k\,r^2/[2\,H_\|\,(1-k\,r^2)]$, the scale factor along the past null cone of the observer at the centre is then given by:
\be
a_\perp(z)=\exp{\left[\int_0^z{\frac{dz'}{(1+z')\,G(z')\,(1+F(z'))}}\bigg|_{\text{nullcone}}\right]}\;.
\label{Eq:avszint}
\ee
Eqs.~(\ref{Eq:avsz}) and (\ref{Eq:avszint}) are general and do not rely on any approximation. Note that the same integral appears in the constraint Eq.~(\ref{glue}).
In the matter dominated era, $G$ is mostly set by the matter profile. Deviations from $-1$ are thus related to gradients in the matter density, so are present in the LTB scenario as well and can typically induce a shift of $\mathcal{O}(\%)$ in the $a_\perp$ vs. $z$ relation.
The $F$-term, however, induces a deviation from the analogous LTB relation. 
To quantify it, we now consider the solution presented in Sec.~\ref{sec:efe}.

In the LTB coordinate frame (namely, the matter frame), we can set $\phi=\dot \phi=0$, which shows that the correction is associated to the temporal variation of the curvature term. This is induced by the inhomogeneous pressure as outlined in Sec.~\ref{sec:onefluid}. 
Combining the third and fourth equations of the system, it's interesting to note that $d \log{\rho}/dt=-H_\parallel-2\,H_\perp-\dot k\,r^2/[2\,(1-k\,r^2)]$. The latter term, which gives the correction in the matter density evolution with respect to the LTB computation, is the same as the correction in the redshift-relations. Using $v_p=a_\|/\sqrt{1-k\,r^2}\,v$, the $F$-term reads:
\be
F=-\frac{2\,a_\perp\,r}{3\,H_\parallel\,\sqrt{1-k\,r^2}}\mu\,v_p \;,
\label{Eq:Fmat}
\ee
which is proportional to the peculiar velocity. This shows once again that it has its origin in the inhomogeneity of the radiation density and the associated pressure-gradient which induces a peculiar velocity in the radiation component with respect to the matter frame.

To compute the redshift of CMB photons, we focus on the radiation frame, where $v_p=\sqrt{1-k\,r^2}/a_\|\,v$, and
\be
F=\frac{a_\perp\,r}{6\,a_\|\,H_\|}\left(H_\perp \frac{\mu'}{\mu}-2\,e^{\phi}\,\rho\,v\right) 
+\frac{3\,H(r_0)-H_\|-2\,H_\perp}{3\,H_\|} \;,
\label{Eq:Frad}
\ee
where $e^\phi=(\mu(r_0)/\mu)^{1/4}$.
We derive an estimate of the size of $F$ by computing the evolution of $\rho$, $\mu$, and $a_{\perp,\|}$ neglecting velocity terms in the system of Sec.~\ref{sec:efe}, and then plugging in the solutions to derive the evolution of the coordinate velocity $v$ (fifth equation). 
We stress that such estimate only gives a rough order of magnitude idea about the shift in the $a_\perp$ vs. $z$ relation given by inhomogeneous radiation.
To properly compute it, it is crucial to correctly estimate the evolutions of size and gradient of the radiation profile, which requires the integration of the full system of equations (strictly speaking, one cannot separate $\mu$ and $v_p$ evolutions). 

\begin{figure}[hb]
\begin{center}
\includegraphics[width=1\columnwidth]{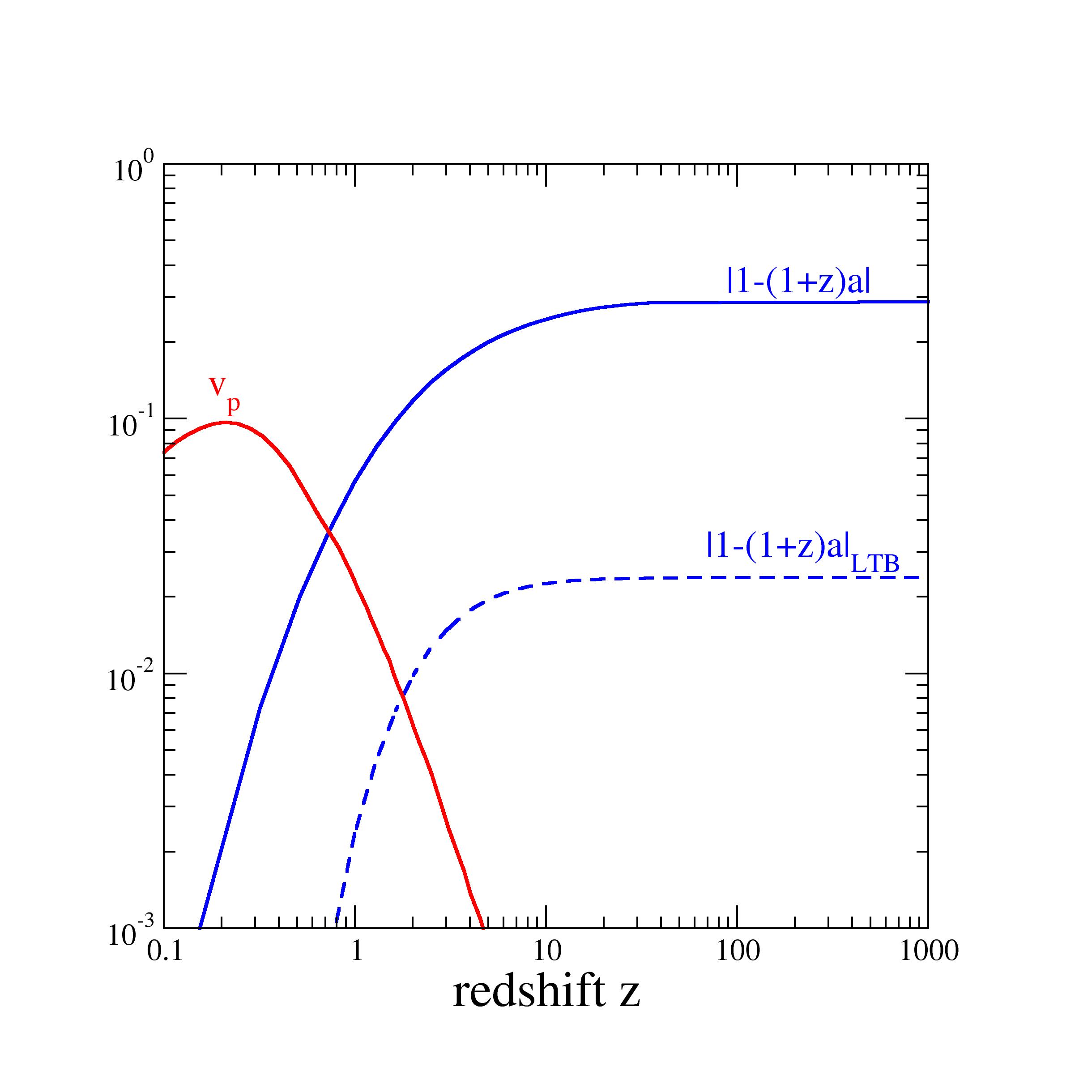}
\caption{Peculiar velocity $v_p$ (red) and $|1-a_\perp\cdot(1+z)|$ (blue) estimated in the approximation described in the text. The deviation of $(1+z)\cdot a_\perp$  from $1$ is computed evaluating Eq.~(\ref{Eq:avszint}) in the radiation frame. The dashed line shows the result neglecting the $F$-term.}
\label{fig:avsz}
\end{center}
\end{figure}

We focus on an example and set parameters as in the benchmark case described in Fig.~\ref{cls} (with the shape of radiation profile the same as the matter).
Results are shown in Fig.~\ref{fig:avsz}.

The main conclusion that can be drawn is that including inhomogeneous radiation $(1+z)\cdot a_\perp$ can significantly differ from $1$ (by $\sim 30\%$ in this approximation) at $z\gg1$. The correction increases corresponding to the profile gradients and then flattens in the asymptotic FLRW region. Critically, we need a shift of just about this amount to satisfy Eq.~(\ref{glue}).

From another point of view this implies that we have a freedom to choose $a_*$ where this choice means that we selected a radiation profile giving the proper shift in Eq.~(\ref{Eq:avszint}). Note that it is analogous to choosing $t_*$ in Eq.~(\ref{glue}).
From the approximated picture in Fig.~\ref{fig:avsz} it looks plausible that a radiation profile with an $\mathcal{O}(1)$ density contrast today (similar to matter) can give $|1-a_*\cdot(1+z_*)|\sim10-20\%$, which is the shift typically required by the fit to CMB data in the models considered in the rest of the paper (see Sec.~\ref{sec:result}) and correspond to $F\sim1\%$. 

Assuming the central wordline to be FLRW, this is provided by an inhomogeneity $\mathcal{O}(1)$ in the radiation density at times around $t_\dec$~\cite{RC}. 
More generally, the study of the class of radiation models selected by $a_*$ requires a numerical solution of the system in Sec.~\ref{sec:efe} (plus initial conditions possibly provided by a production mechanism for the inhomogeneity linking matter and radiation initial profiles) and deserves a dedicated work.
We will consider $a_\dec^\out$ as a free parameter (set by $T_0^\out$, as explained above).

Similarly, a correction is expected also in the computation of the coordinate distance $r$ from Eq.~\ref{dzdr}. The difference $|1-r/r_{LTB}|$ can be sizable ($\mathcal{O}(10\%)$ for profiles and approximation illustrated here) and the shift goes in the opposite direction with respect to the shift in $a_*$ discussed above. On the other hand, to precisely state whether or not they cancel in the area distance $d_A=a_*r_*$  requires the integration of the system in Sec.~\ref{sec:efe} and depends on the details of the radiation profile. It is possible that they don't, in which case some of our results in Sec.~\ref{sec:da} will change; in particular, it is conceivable that an asymptotically flat model may be possible.
We will simply assume $d_A$ to follow from LTB equations, which is our main approximation (or, alternatively, it can be seen as a restriction of the analysis to a particular class of radiation profiles).

Note that, in order to estimate $F$, we need to compute the peculiar velocity. 
Our results roughly compare to Refs.~\cite{Alnes:2006pf,Caldwell:2007yu,GarciaBellido:2008gd}, where the velocity is computed in an LTB scenario considering the dominant contribution given by the dipole, i.e., $v_p/c=(\Delta T/T)_{dipole}$, and to Ref.~\cite{Garfinkle:2009uf}, where a covariant formalism is introduced.
However, our result is strongly dependent on the assumption for the radiation profile (and on the approximation considered for the solution).
Therefore we stress that in order to have a complete estimate of the peculiar velocity, and in turn of the kinematic Sunyaev-Zeldovich effect in void models, one has to take into account the effect of inhomogeneous radiation in geodesic equations. It will probably lead kSZ data to select radiation profiles rather than rule out matter inhomogeneity, as discussed in Sec.~\ref{sec:disc}. 

\subsection{What if the central worldline is not FLRW?}

While we have argued that the central worldline can be considered to be FLRW to a good enough approximation (rather, we used this assumption to calculate early-time conditions for matter and radiation; it is not directly used to calculate constraints from the CMB), let us assume for a moment that this is incorrect. The radiation temperature at the centre is determined by the radiation streaming into the central worldline along null-cones, and so one can argue that it may be determined by the redshift to a surface of constant time in an LTB model, as opposed to the evolution of radiation along the central worldline.  (Of course, if the full spacetime solution were known these would be the same.) As argued above, we need to know precisely the surface $t_*(T_*)$ to calculate this accurately, but the LTB approximation at least illustrates the key idea for this central temperature calculation. 
Let us consider where the central worldline evolution changes our analysis. 

In the radiation frame, the temperature always behaves as $T\propto1/a_\mathrm{rad}$, where $a_\mathrm{rad}$ is the mean length scale in that frame (defined covariantly through the frame expansion rate). In the matter frame the temperature evolution is more complicated because of the radiation flowing through the frame. At the centre, let us assume that it can be calculated at any time $t$ by $T=(1+z_\#)T_\#$, where \# denotes the surface of constant time $t_\#$, in a pure dust LTB model. If we do this we find, approximately, 
\be
T=\left\{
\begin{array}{ccc} \displaystyle
  \frac{T_0}{a^n}   &  \mathrm{for} & a>\hat a \\[2mm]
  \displaystyle  \frac{T_0 \hat a^{1-n}}{a}   &  \mathrm{for} & a<\hat a ,
\end{array}
\right.
\ee 
where $\hat a$ is the value of the scale factor (at the centre) where the transition from normal $1/a$ scaling changes (i.e., when the central observer starts to see the void), and $n$ gives the scaling behaviour at late times ($n=1$ corresponds to FLRW evolution). 
The consequences of this are that:
\begin{itemize}
\item The baryon-photon ratio evolves: we have $\eta_0^\in\simeq \hat a^{3(1-n)}\eta_\dec^\in$.
\item The age calculated at the centre will be different.
\item $T_\dec^\in$ and $T^\in_\eq$ will change.
\end{itemize}
So, if we take $^7$Li constraints on $\eta^\in$, for example, then this is the early time value, giving $\eta_\dec^\in$, and is less today by about a factor of two or so (depending strongly on the void parameters), given by $\hat a^{3(1-n)}$.  This either requires a lower $h^\in$ or $\omin$ if $f_b=\,$const., or a lower $f_b$ at the centre~-- see Eq.~(\ref{varpi_b}). In fact, if we enforce $f_b=\,$const. and use a very low value of $\eta^\in_0$, as estimated from this LTB scaling law, we find models similar to~\cite{ZMS,CFZ,BNV,MZS}. As far as the age is concerned, the change is tiny since radiation is subdominant when the change in scaling takes place (changing, e.g., $\omin$ by a fraction of a percent would adjust for this). The rest of our analysis goes through as discussed, and, in particular, the constraint Eq.~(\ref{glue}) will be automatically satisfied. The fact that $T_\dec^\in$ and $T^\in_\eq$ change makes no difference to our analysis; these are determined after-the-fact anyway, as illustrated in Fig.~\ref{overview}. Of course, as we have discussed in this Appendix, we are unable to calculate $T$ along the central worldline in the real model following this approach because we don't accurately know where $t_*$ is when we integrate from the central worldline down a past lightcone. Finally, we note that specifying a temperature evolution law for the central worldline is akin to choosing an inhomogeneous radiation profile: it is clear from this consideration too that we are free to do just this.

\end{document}